\documentclass[final,3p,12pt]{elsarticle}

\pdfoutput=1

\usepackage{graphicx}

\usepackage{amssymb}

\journal{Surface Science}

\begin{document}

\begin{frontmatter}



\title{Computational investigation of the temperature influence on the cleavage of a graphite surface}


\author{N.V.~Prodanov\corref{cor}}
\ead{prodk@rambler.ru}

\author{A.V.~Khomenko}
\ead{khom@mss.sumdu.edu.ua}
\ead[url]{http://personal.sumdu.edu.ua/khomenko/eng/}

\cortext[cor]{Corresponding author: Tel.: +38 0542 333155; Fax: +38 0542 334058 }

\address{Sumy State University, 2 Rimskii-Korsakov Str., 40007 Sumy, Ukraine}

\begin{abstract}
Mechanical exfoliation of a graphite surface with an adhesive nanoasperity is studied under different temperatures ranging from 298~K to 2~K using classical molecular dynamics. Two types of the interlayer interaction are investigated. For a pairwise Lennard-Jones potential the complete removal of the upper graphene layer during the retraction of the nanoasperity occurs in the whole range of the temperatures considered. The results obtained using registry dependent potential, which takes into account electronic delocalization contribution besides the van der Waals one, exhibit more pronounced temperature dependence. In this case the exfoliation takes place for temperatures higher than 16~K, but beginning from 8~K down to 2~K the system behavior manifests qualitative changes with the absence of cleavage of the sample. Analytical estimates combined with the results of the simulations reveal that the contribution of the overlap of $\pi$ orbitals of carbon atoms plays an important role in the exfoliation of graphite.
\end{abstract}


\end{frontmatter}


\section{Introduction}
\label{intro}

Graphite is a lamellar material which is widely used in the experiments where an atomically flat surface is required and its fabrication is accomplished by mechanical cleavage of a graphite sample~\cite{Diet2008,Hols2008,Dien2004,DienThes}. The cleavage of graphite is usually considered in the mentioned applied context. However, understanding the detailed physics of this process and elucidating the influence of different factors on its behavior may be valuable both from practical and theoretical  viewpoints. In this context it is worth mentioning that mechanical exfoliation was the technique which allowed the discovery of graphene~\cite{Novos2004}, a monolayer of carbon atoms tightly packed into a honeycomb lattice. This novel material has unusual electronic properties and it is promising for a wide variety of applications, in particular, the creation of new high-frequency electronic devices~\cite{Geim2007,Castr2009}. In spite of the development of new methods for producing graphene at high yields~\cite{Herna2008,Qian2009,Geim2009}, micromechanical cleavage or exfoliation of bulk graphite still remains the main technique used by most experimental groups for the fabrication of high-quality graphene samples~\cite{Geim2009,Nemes2008,Elias2009}. Comprehensive understanding of this process may assist in adjusting the conditions for production of samples with the desired characteristics. The cleavage of graphite is also closely related to the so-called superlubricity which was observed during probing a graphite surface with tungsten tip of a friction force microscope (FFM)~\cite{Dien2004,DienThes}. This phenomenon is characterized by a reduction of friction by orders of magnitude and it is attributed to the existence of a small graphite flake attached to the tip~\cite{Dien2004,DienThes,Dien2004B,Fili2008}. Revealing the contributions of different factors to the formation of a graphitic flake, which occurs by cleavage from a graphite surface, may be valuable for the establishing the conditions of realization of the superlubricity phenomenon.

In spite of practical significance of the graphite exfoliation there is a lack of its theoretical studies. Models of superlow friction of graphite are often based on the assumption of the presence of the cleaved graphitic layers~\cite{DienThes,Dien2004B,Fili2008,Mats2005}. There are also theoretical investigations of nanoindentation of graphite using classical molecular dynamics (MD)~\cite{Harri1992,Richt2000,Guo2004} or boundary element method~\cite{Yang2009}. Diamond~\cite{Harri1992} and virtual indentors~\cite{Richt2000,Guo2004,Yang2009} are employed to probe the mechanical properties of graphite~\cite{Harri1992,Richt2000,Yang2009} or to explore the formation of interlayer $sp^{3}$ bonds under high pressures~\cite{Guo2004}. However, repulsive interactions between the indentor and the sample in the works mentioned above do not allow the investigation of mechanical exfoliation of graphite which could have been observed for the adhesive tips. Some theoretical analysis of graphite cleavage can be found in Ref.~\cite{Liang2009} where novel fabrication method for incorporating nanometer to micrometer scale few-layer graphene features onto substrates with electrostatic exfoliation is described. Numerical simulations represented in Ref.~\cite{Liang2009}, however, are intended only to determine the field strengths needed for performing the described process and do not reveal the accompanying physics.

To fill up the gap in theoretical studies of graphite cleavage we have carried out computer experiments using classical MD. The considered model resembles ones described above for the graphite indentation, but it has the following two principal differences. The first one is the use of the adhesive tip. Note, that the indentation, which occurs as a consequence of a jump-to-contact, is very shallow and is not the target of the current investigation. The second difference pertains to the interlayer interaction. In the mentioned works~\cite{Harri1992,Richt2000,Guo2004} a pairwise Lennard-Jones potential (LJP) which takes into account only van der Waals (vdW) attraction between the layers is used. However, as studies exploiting quantum-mechanical techniques suggest, there is also a short-ranged electronic delocalization contribution to the interlayer bonding of graphite~\cite{Charl1994,Kolmo2005}, and neglecting it may influence the exfoliation process. Thus, we performed the separate simulations using the LJ and the registry dependent potential (RDP)~\cite{Kolmo2005}, which includes the mentioned orbital overlap contribution. The main aim of the work is to analyze the graphite cleavage under different temperatures using these two interlayer potentials. Temperature of the sample is one of the natural factors that has an impact on the interlayer cohesion in graphite, and it has been recently used in the solvothermal-assisted method of graphene production~\cite{Qian2009}, hence indicating the need of its thorough exploration. The next section gives the details of the simulation setup.

\section{Model}
\label{model}

\begin{figure*}[htb]
\centerline{\includegraphics[width=0.48\textwidth]{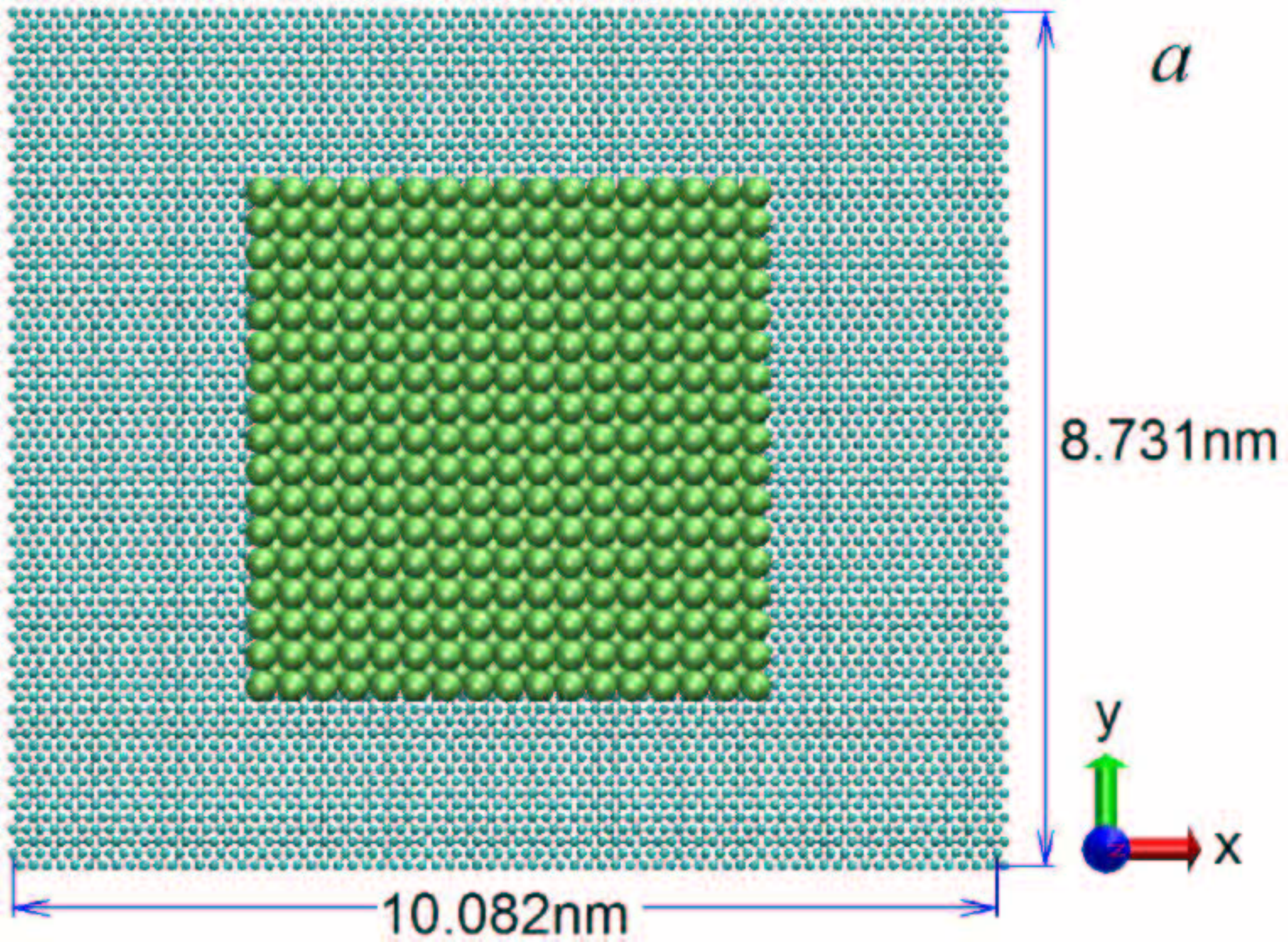}
\includegraphics[width=0.52\textwidth]{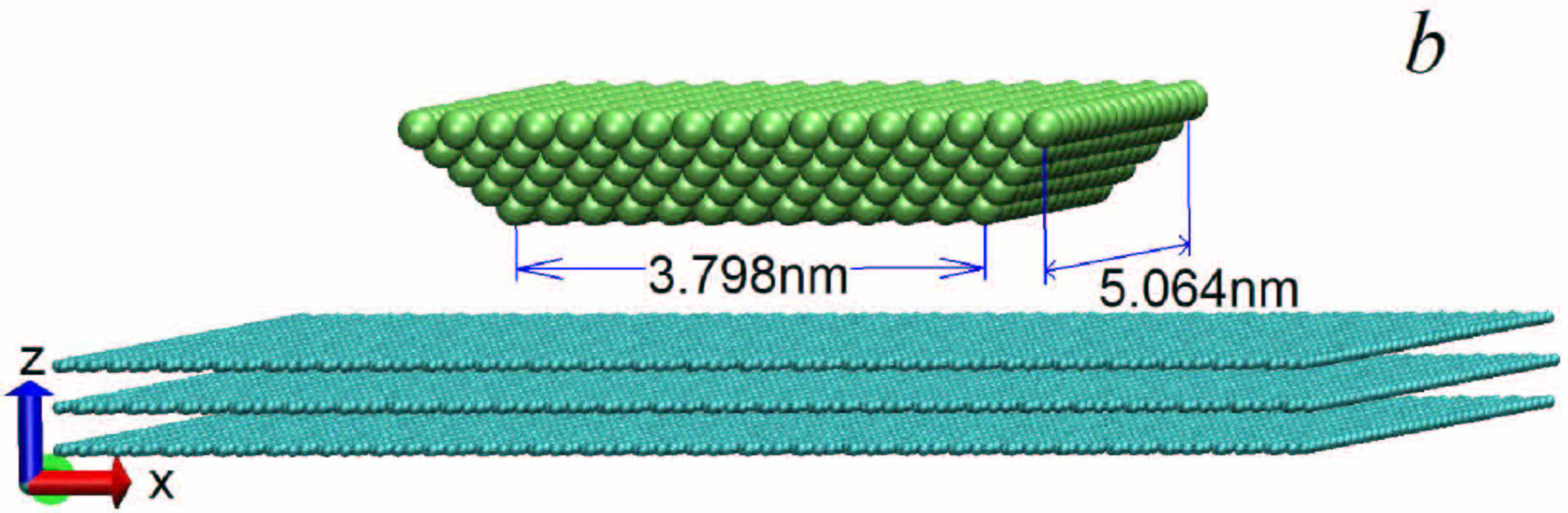}}
\caption{Top (\textit{a}) and perspective (\textit{b}) views of the initial atomic configuration of the studied system. Green and cyan balls correspond to tungsten and carbon atoms respectively (all snapshots in this work are produced with Visual Molecular Dynamics software~\cite{vmd}).}
\label{fig1}
\end{figure*}

The graphitic sample consists of three graphene layers with AB stacking~(fig.~\ref{fig1}) which reflects $\alpha$ form of graphite. Armchair and zigzag graphene edges lie along $x$ and $y$ coordinate axes, respectively, and periodical boundary conditions are applied in the $xy$-plane. Each layer is composed of $24\times24$ honeycombs thus containing 3456 carbon atoms and the lengths along $x$ and $y$ directions are 10.082~nm and 8.731~nm, respectively. To hold the sample in space, the bottom graphitic layer is rigid throughout the simulations.

Our model is mainly approached to the experiments pertaining to the superlubricity. The graphite surface interacts with an infinitely hard square pyramidal nanoasperity (to which we also refer as the tip) which simulates the tip of FFM. The asperity consists of five layers of atoms parallel to the $xy$-plane. Particles are arranged in a perfect bcc lattice with constant of 0.3165~nm and this corresponds to the crystal structure of tungsten~\cite{webel}. The tapered form is provided by adding one atomic row in $x$ and $y$ directions per layer when moving from the bottom (which is the nearest to the sample part of the asperity) to the top of the tip. The bottom atomic layer exposes (001) crystallographic plane and has $13\times13$ atoms on the area~$a_{x}\times a_{y}$. The nanoasperity contains 1135 atoms and the total number of particles involved in the simulations is 11503.

It should be noted, that the hardness of the nanoasperity may influence the exfoliation and, strictly speaking, for completely realistic reproduction of the experimental conditions the tip in the model should be able to deform. This can be achieved by exploiting one of the available interatomic potentials for tungsten, e. g. based on the modified embedded atom method~\cite{Baske1992} or its new form~\cite{Lee2001}. Nevertheless, absolutely rigid surfaces are quite often used in MD simulations~\cite{Koman2000,Khome2008} and we also decided to study the system under mentioned approximation as the first step towards more realistic modeling.

Nanoasperity dimensions are chosen to satisfy the fact that accordingly to the experiments the flake is assumed to attach to asperities on the tip with sizes of several nanometers (see high-resolution transmission electron microscopy micrograph of the tungsten tip in fig.5.11 in Ref.~\cite{DienThes}). The size of the nanoasperity greatly affects the exfoliation in our simulations (this is analyzed in section~\ref{discussion}) and its value has been chosen to provide the most suitable conditions for the demonstration of the differences in system behavior with LJP and RDP.

Covalent bonds between carbon atoms within the two upper dynamic graphene layers are described by the Brenner potential. It has the following form~\cite{Brenn1990,REBO2002}
\begin{equation}
\label{eq1-brenner}
V_{B}=\sum_{i}\sum_{j>i}
[V^{R}(r_{ij}) - \overline{b}_{ij}V^{A}(r_{ij})].
\end{equation}
In the current study expressions of a second-generation reactive empirical bond order (REBO2) form of the potential~\cite{REBO2002} are used for pair-additive interactions $V^{R}(r_{ij})$ and $V^{A}(r_{ij})$. Bond order function $\overline{b}_{ij}$ is chosen as in the first version of the Brenner potential (REBO1) with parameters for potential II in Ref.~\cite{Brenn1990}. The code from TREMOLO software~\cite{Grieb2007} is partly used in calculations of cubic splines and their derivatives in the bond order term, and the interactions from Brenner potential are computed using parallel algorithm presented in Ref.~\cite{Cagla1999}.

The use of pairwise interactions from REBO2 in the current model is caused by the fact that REBO1 is incapable of proper description of any short-range hard wall repulsion that might prove important under high compression~\cite{REBO2002,Harri1991}. However, more complex form of the bond order term in REBO2 and hence more intensive computations forced us to use the $\overline{b}_{ij}$ from REBO1 because of the computational restrictions. There are, however, several arguments that justify its use in the context of our problem. First, let us analyze the roles that different contributions play in the REBO potential. The energetics of each given hydrocarbon structure is defined by the pairwise terms $V^{R}(r_{ij})$ and $V^{A}(r_{ij})$ with the latter modulated by the bond order function $\overline{b}_{ij}$. The main aim of $\overline{b}_{ij}$ is to appropriately adjust the energy of the atomic structure when the changes in the local atomic environment occur. This is accomplished by tracking the number of the nearest carbons and hydrogens and the angle $\theta$ between the neighboring atoms. If the atomic structure is not changed, the universal function $\overline{b}_{ij}$ can be, roughly speaking, superseded by its numerical value for the current configuration. In our model the atomic coordination is not assumed to alter considerably, so we could have used the corresponding numerical value instead of $\overline{b}_{ij}$ in eq.~(\ref{eq1-brenner}). But to take into account the changes in the local environment due to thermal fluctuations we have exploited the function $\overline{b}_{ij}$ from REBO1, which gives the value of $\overline{b}_{ij}$ close to the one from REBO2 for graphite at the equilibrium. Substituting $\theta=120^{\circ}$ and $r=1.42$~\AA~in expressions for $\mathrm{G}(\theta)$ and $V_{B}$ for REBO1~\cite{Brenn1990} one obtains $\mathrm{G}(\theta)=0.0372$, $\overline{b}_{ij}=0.9648$ and $V_{B}=-5.2854$~eV. For REBO2 the corresponding values are~\cite{REBO2002}: $\mathrm{G}(\cos(\theta))=0.0528$, $\overline{b}_{ij}=0.9511$ and $V_{B}=-4.9861$~eV, indicating the difference in binding energy between the two potentials only in 6~\%. This is admissible for our problem because taking into account other approximations of the model we do not pretend to obtain accurate quantitative results. Note, however, that considerable deviation of an atom from the equilibrium position may lead to an instability as the consequence of inability of the used $\overline{b}_{ij}$ to appropriately describe considerable changes in atomic coordination, resulting in the rearrangements of atoms and the formation of defects. As the simulations show, in our case this can be observed for temperatures higher about 25~K (see section~\ref{discussion}). Nevertheless, the main results of our computer experiments pertain to lower temperatures where the instabilities do not occur and hence the overall conclusions of the study are not altered.

As another arguments for the use of such a potential it should be noted that approximations for in-plane interactions in graphite are often employed when the main concern is directed at the interlayer processes~\cite{DienThes,Dien2004B,Fili2008,Mats2005}. Speaking about inaccurate stiffness of the layers, as our estimates in section~\ref{discussion} show, it might have quantitative but not qualitative influence on the exfoliation and, in addition, our model provides the prospects for the investigation of the effect of stiffness of layers on the considered process. Lastly, a uniform use of just one in-plane interaction allows comparing the behavior obtained with different interlayer potentials, which is the aim of the present study.

For modeling of the cleavage of graphite the crucial role may play the proper description of the interlayer binding~\cite{Kolmo2005,Ito2008}. A pairwise Lennard-Jones potential (LJP) can describe the overall cohesion between graphene layers, but it is much too smooth to reflect variations in the relative alignment of adjacent graphitic interfaces, which is also true for other graphite potentials, such as AIREBO~\cite{Stuar2000}, that use LJ interaction for the coupling of layers. The reason for this is that the corrugation is mainly defined by the electronic delocalization contribution~\cite{Charl1994,Kolmo2005}, which is anisotropic and cannot be described in a natural way by the single length scale of a Lennard-Jones potential. Since during the exfoliation the upper graphene layer is being deformed, the anisotropic interaction may be significantly changed, which might greatly affect the considered process. To reveal this effect we explore two types of the interlayer potential. The first is a registry-dependent interlayer potential (RDP) that has the following form~\cite{Kolmo2005}:
\begin{eqnarray}
\label{eq2-kolmogorov}
V(\mathbf{r}_{ij},\mathbf{n}_{i},\mathbf{n}_{j})=
e^{-\lambda(r_{ij}-z_{0})}[C+f(\rho_{ij})+f(\rho_{ji})]
\nonumber\\
-A\left(\frac{r_{ij}}{z_{0}}\right)^{-6},
\end{eqnarray}
where $\lambda=3.629$~\AA$^{-1}$, $z_{0}=3.34$~\AA, $C=3.030$~meV, $A=10.238$~meV. The potential contains an $r^{-6}$ vdW attraction and an exponentially decaying repulsion due to the interlayer wave-function overlap.
To reflect the directionality of the overlap the function $f$ is introduced which rapidly decays with the transverse distance $\rho$. The latter is defined via the distance $\mathbf{r}_{ij}$ between pairs of atoms $i$ and $j$ belonging to distinct layers and the vector $\mathbf{n}_{k}$ ($k=i,j$) which is normal to the $sp^{2}$ plane in the vicinity of atom $k$. In the present study $\mathbf{n}_{k}$ is computed as ``local'' normal, i. e. as average of the three normalized cross products of the displacement vectors to the nearest neighbors of atom $k$, and this corresponds to RDP1 in Ref.~\cite{Kolmo2005} (see section~\ref{discussion} for more details). For long-range vdW term the cutoff distance is equal to $r_{\mathrm{c}}=2.7z_{0}=0.9018$~nm. The presence of normals in the RDP makes it in essence a many-body potential which requires much more computational effort as compared to simple pairwise potentials. In the current study interactions only between the adjacent layers are considered and they are computed using our specially developed parallel algorithm based on linked cell lists~\cite{Grieb2007,Rapap2004}.

In the second series of the simulations the interlayer energy is represented by pairwise LJP:
\begin{equation}
\label{eq3-LJ-carbon}
V_{LJ}=\left\{\begin{array}{lr}
         4\varepsilon_{CC}\left[
         \left(\frac{\sigma_{CC}}{r}\right)^{12}-
         \left(\frac{\sigma_{CC}}{r}\right)^{6}\right], & r < r_{\mathrm{c}}\\
         0, & r \geq r_{\mathrm{c}}
       \end{array}\right.,
\end{equation}
where $r$ is the distance between a pair of carbon atoms, values of $\varepsilon_{CC}=2.8$~meV and $\sigma_{CC}=3.33$~\AA~were adjusted to obtain the interlayer binding energies and spacings close to ones that RDP gives, and the cutoff distance $r_{\mathrm{c}}$ is the same as for the RDP.

The tip is assumed to interact only with the upper graphitic layer. LJP with parameters $\varepsilon_{WC}=0.5$~eV, $\sigma_{WC}=0.5z_{0}$ corresponding to $\varepsilon_{CC}$, $\sigma_{CC}$ in eq.~(\ref{eq3-LJ-carbon}), is used for interactions between the tungsten and carbon atoms. These values of $\varepsilon_{WC}$ and $\sigma_{WC}$ provide the conditions for the exfoliation of the upper graphene layer under room temperature in our model~\cite{Khome2010}. The equations of motion are integrated using the leapfrog method~\cite{Rapap2004} with a time step $\Delta t=0.1$~fs.

The necessary temperature is maintained through the Berendsen thermostat coupled with two dynamic graphitic layers and implemented as in Ref.~\cite{Grieb2007}. The thermostat accounts for the numerical and round errors accumulating at each integration step, and also provides the means to remove excess heat, which occurs as the result of work done on the system during approach and retraction of the tip. The implementation of the thermostat includes velocity rescaling by the factor $\beta$ of the following form:
\begin{equation}
\label{eq-beta}
\beta=\sqrt{1+\gamma(\frac{T_{0}}{T}-1)},
\end{equation}
where $T_{0}$ and $T$ are the desired and current temperatures of the system, respectively, and $\gamma\in[0;1]$ characterizes the rate of heat dissipation. In our simulations $\gamma=0.4$ is used, which corresponds to rather strong coupling to the heat bath in order to prevent the destruction of the upper graphene layer. It should be noted, that velocities are rescaled not every time step, but every ten time steps (or 1~fs in physical units), and in the intermediate steps the system is integrated without scaling. This allows to reduce the effect of the velocity scaling on the distribution of energy in the system. For more details see~\cite{Grieb2007,HeoSJ2005}.

The movement of the tip proceeds as follows. After equilibration of the system during 1~ps with the tip outside the range of interaction hung at 1.16~nm above the surface, the asperity was lowered towards the sample. Motion of the tip occurs by changing $z$-coordinates of tungsten atoms in increments of 0.0106~nm. The entire system is equilibrated for 40~fs in between displacements of the nanoasperity. After reaching the minimum distance between the proximal atomic layers of the two interfacing materials of about 0.108~nm, the tip is immediately pulled away from the surface. Mentioned quantities correspond to the rate of the tip movement equal to 265~m/s. The duration of the simulations is 10 or 13~ps.

\section{Results}
\label{results}

During 1~ps after the equilibration period at $T=298$~K, when the forces between the tip and the sample are still zero, the average values of the interlayer distance and the binding energy of the upper two dynamic graphene layers are about $0.336\pm0.003$~nm and $42\pm1$~meV, respectively. These values differ from 0.334~nm and 48~meV computed for rigid layers using RDP~\cite{Kolmo2005} by about 1~\% and 15~\%. The discrepancy may be attributed to the finite cutoff distance used in the present study, thermal fluctuations of normals and the use of local normals instead of semilocal ones. Nevertheless, obtained quantities are very close to the experimental values~\cite{Kolmo2005}. For the LJ interaction the corresponding values at $T=298$~K are $0.337\pm0.003$~nm and $41.8\pm0.4$~meV.

Let us consider the results obtained with RDP and LJP separately.

\subsection{Exfoliation using RDP}
\label{RDP}

\begin{figure*}[!]
\centerline{\includegraphics[width=0.51\textwidth]{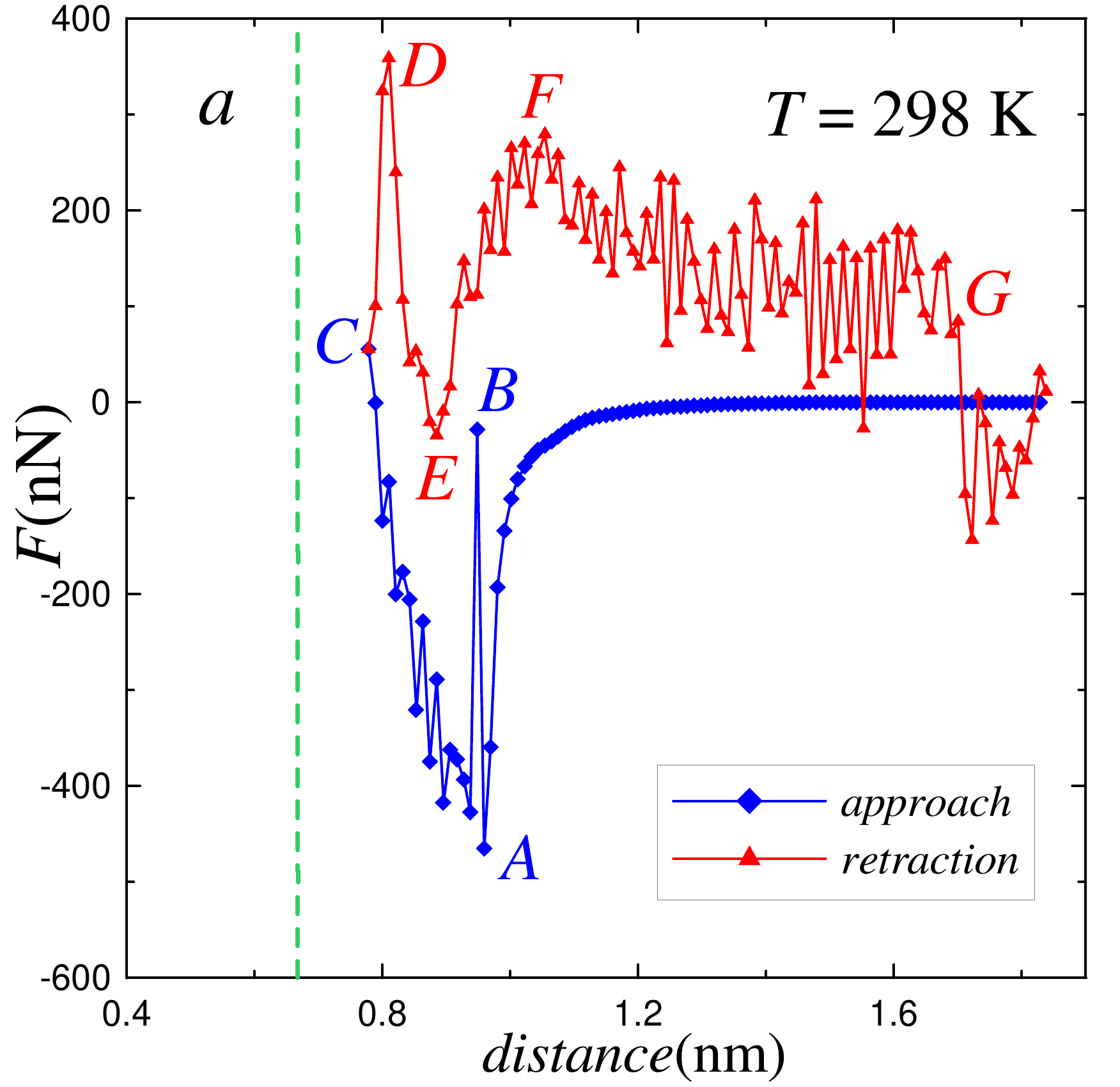}
\includegraphics[width=0.51\textwidth]{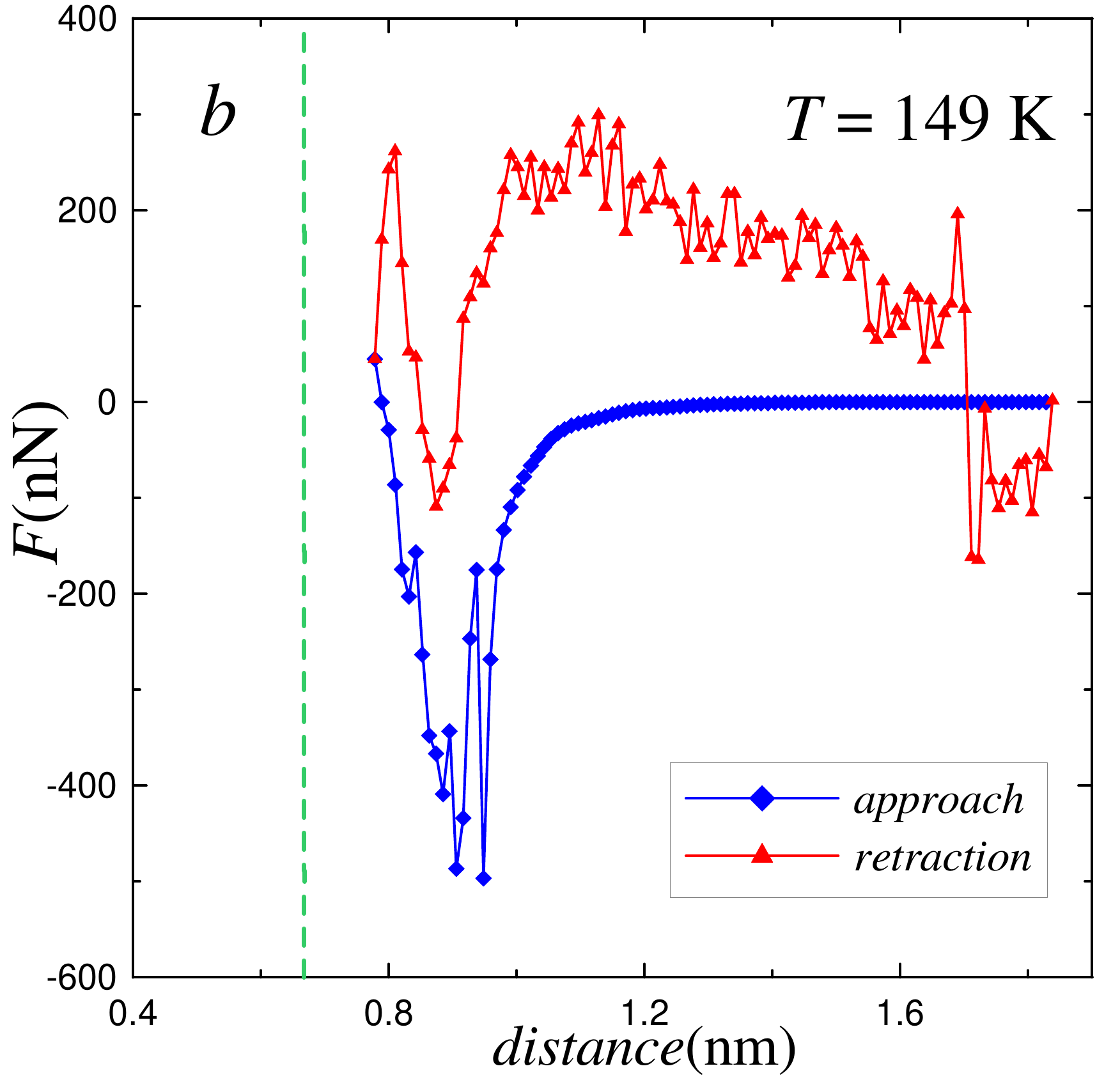}}
\centerline{\includegraphics[width=0.51\textwidth]{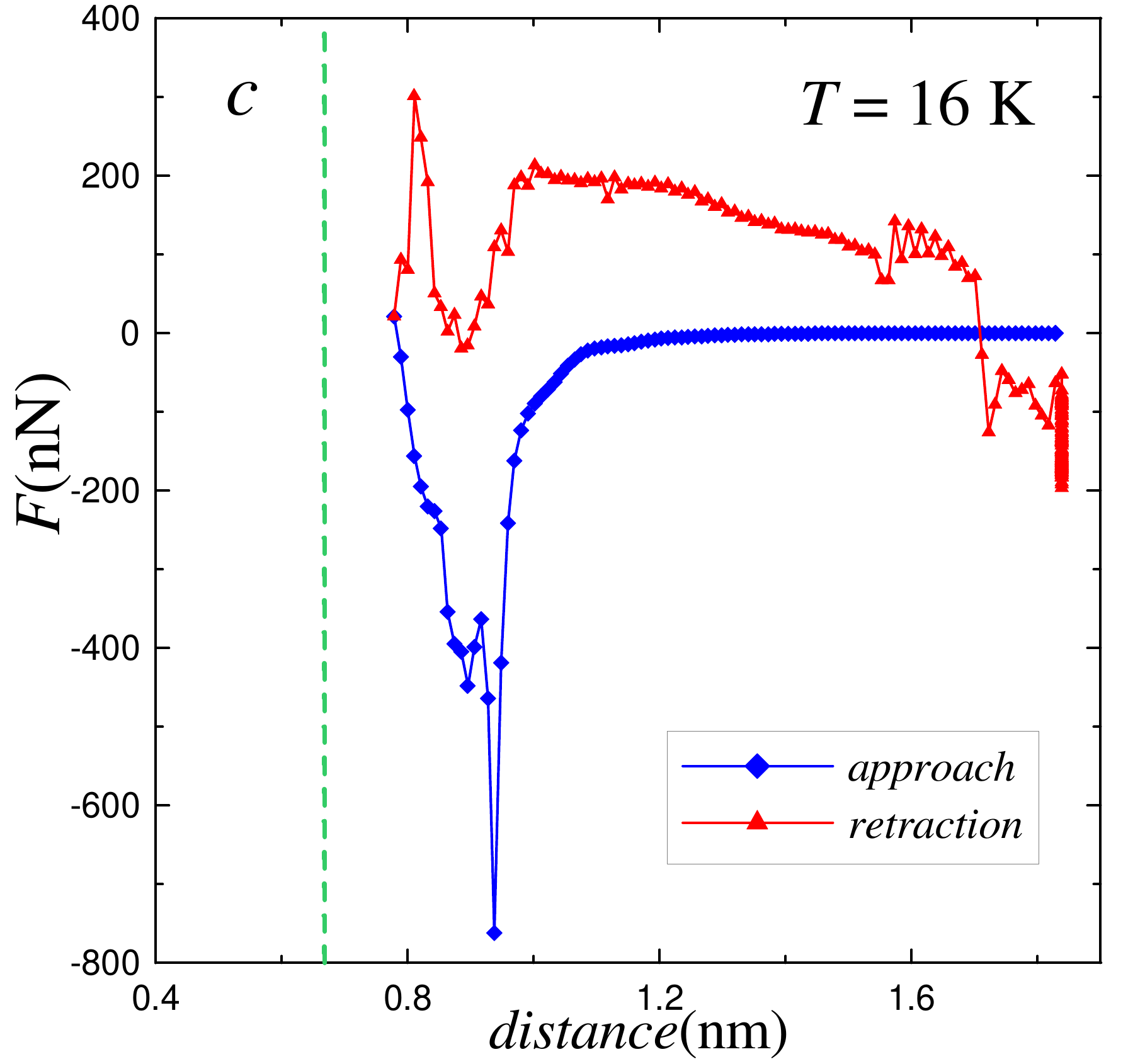}
\includegraphics[width=0.51\textwidth]{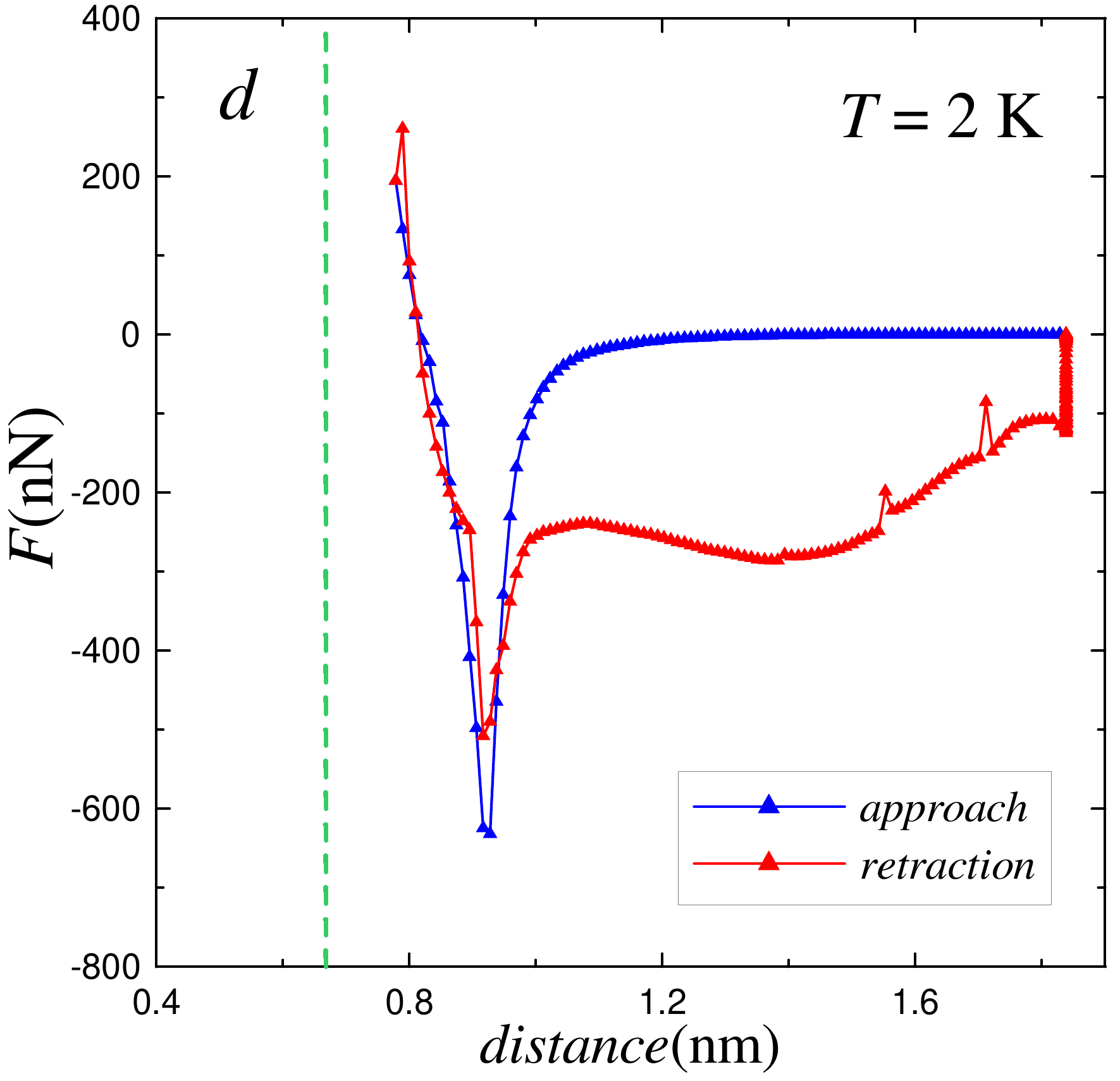}}
\caption{Normal force acting on the nanoasperity as it moves towards and then withdraws from the graphitic surface. Abscissa values correspond to the vertical distance between the rigid graphene layer and the bottom tungsten atomic layer. Dashed line presents the average equilibrium position of the upper carbon layer which is assumed to be 0.668~nm from the bottom carbon layer.}
\label{fig2}
\end{figure*}

Figure~\ref{fig2} shows force-versus-distance curves for different temperatures $T$ obtained when RDP is used. These plots reflect the changes in the normal force $F$ acting on the tip with a distance to the surface. In the present work this force is computed as the sum of $z$ components of forces acting on tungsten atoms from the graphitic sample, and they are averaged over the last 10~fs of the equilibration procedure in between displacements of the tip. In some works additional averaging is performed on the force-displacement curves to filter out the noise from thermal vibrations~\cite{Garg1998,Garg1999}. This has not been carried out in the present study.

\begin{figure*}[!]
\centerline{\includegraphics[width=0.51\textwidth]{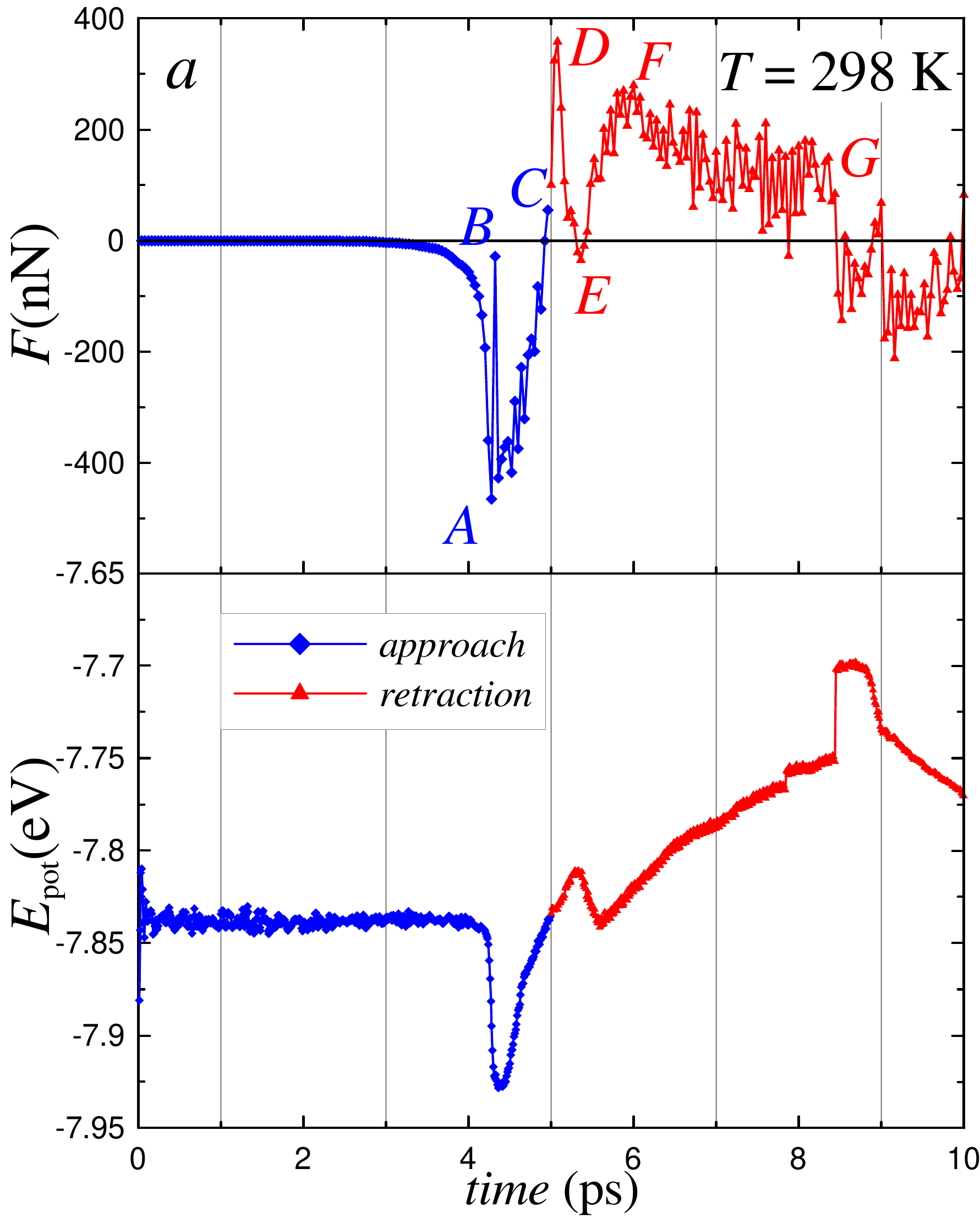}
\includegraphics[width=0.51\textwidth]{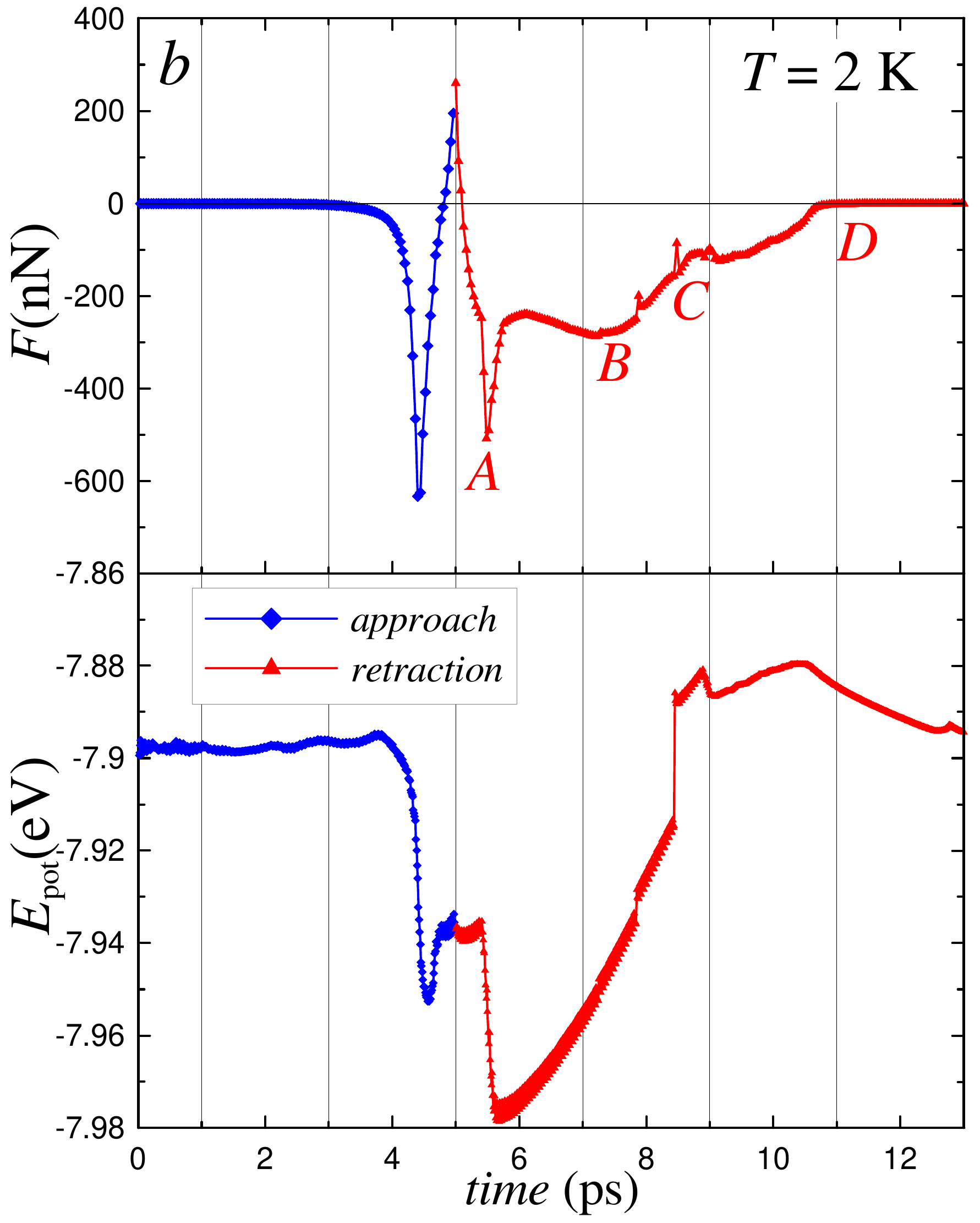}}
\caption{Time dependencies of the normal force acting on the tip and the potential energy of the system (per atom) for the maximum (\textit{a}) and minimum (\textit{b}) temperatures used in the simulations.}
\label{fig3}
\end{figure*}

\begin{figure*}[htb]
\centerline{\includegraphics[width=0.51\textwidth]{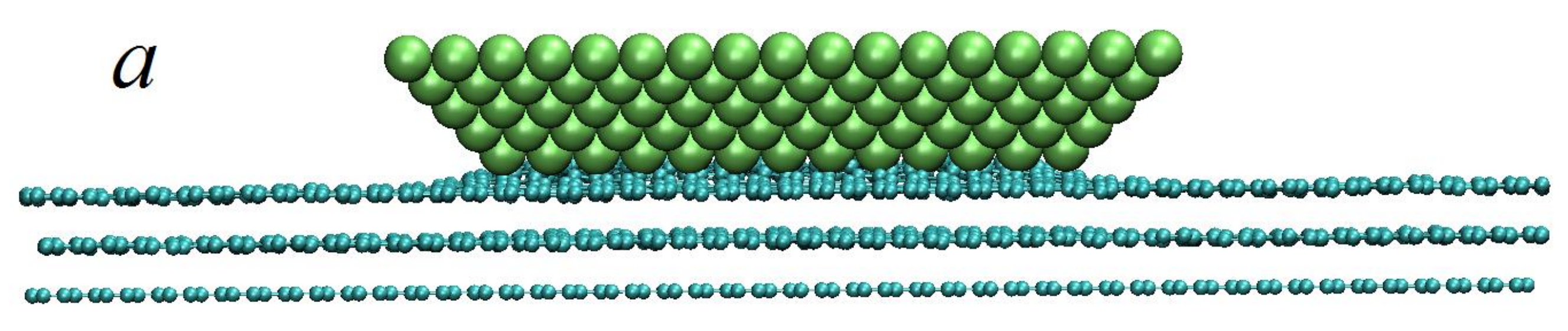}
\includegraphics[width=0.51\textwidth]{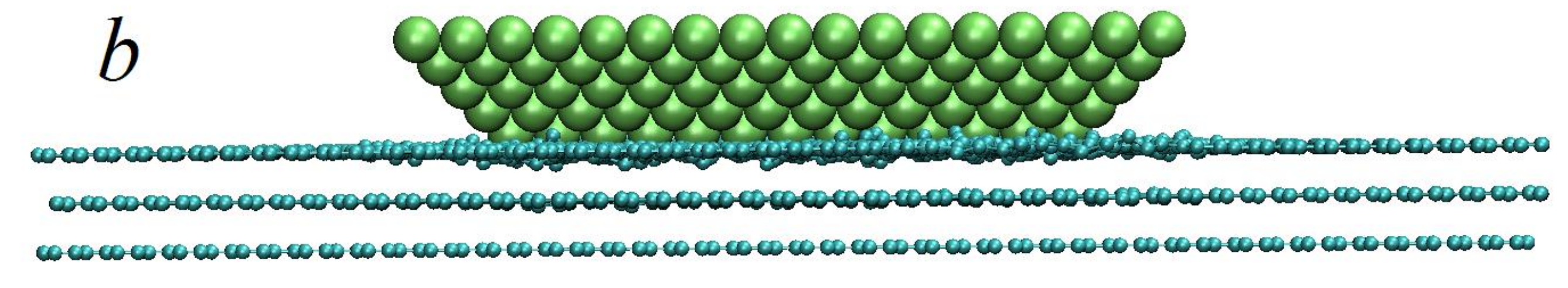}}
\centerline{\includegraphics[width=0.51\textwidth]{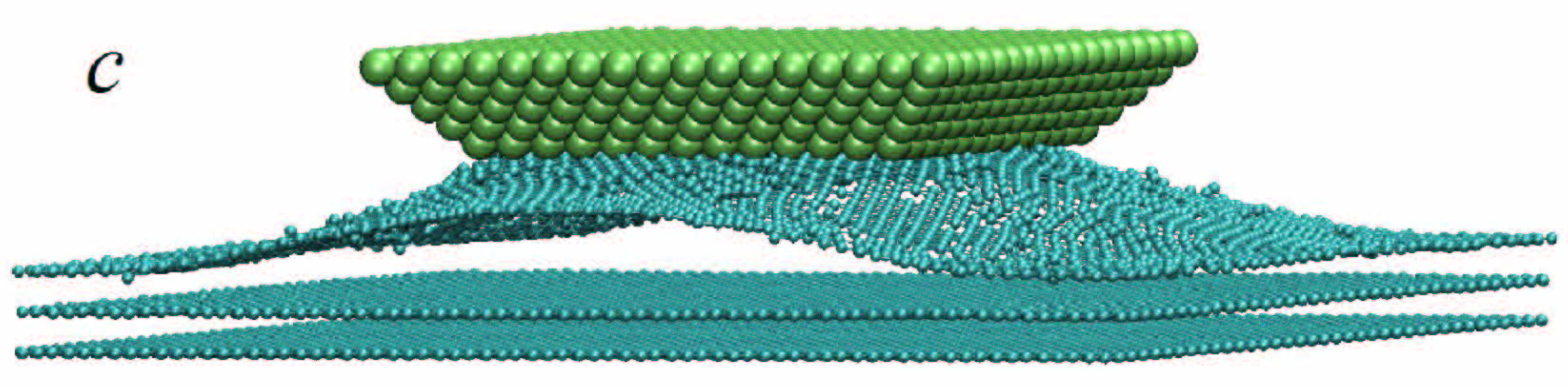}
\includegraphics[width=0.51\textwidth]{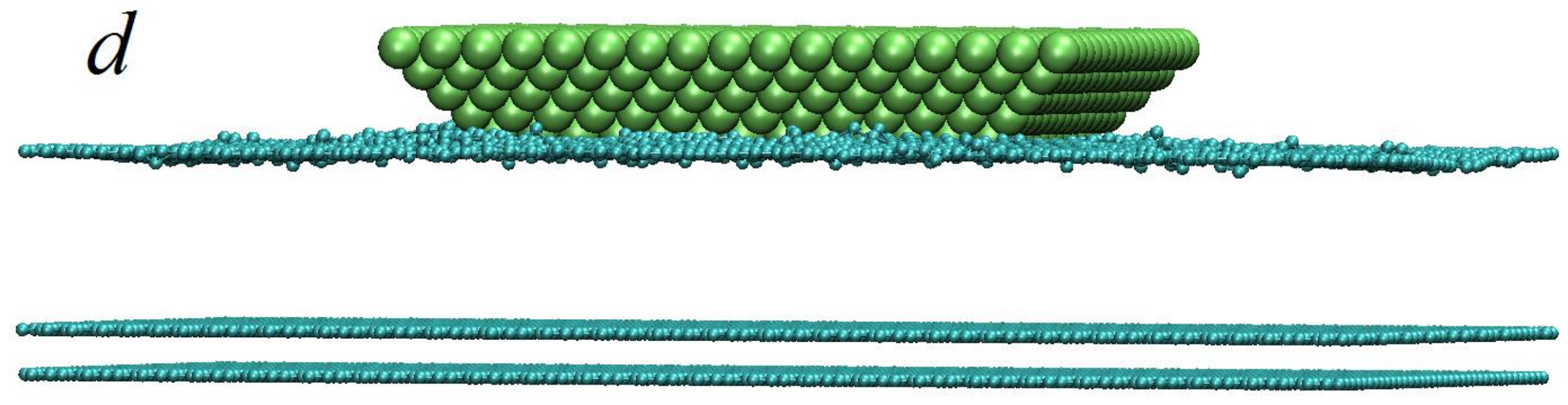}}
\caption{Instantaneous atomic configurations when $T=298$~K corresponding to the following points in fig.~\ref{fig2}\textit{a} and fig.~\ref{fig3}\textit{a}: (\textit{a}) point \textit{A}, (\textit{b}) \textit{D}, (\textit{c}) \textit{G}, (\textit{d}) exfoliation of the upper layer at the end of the simulation.}
\label{fig4}
\end{figure*}

\begin{figure}[htb]
\centerline{\includegraphics[width=0.52\textwidth]{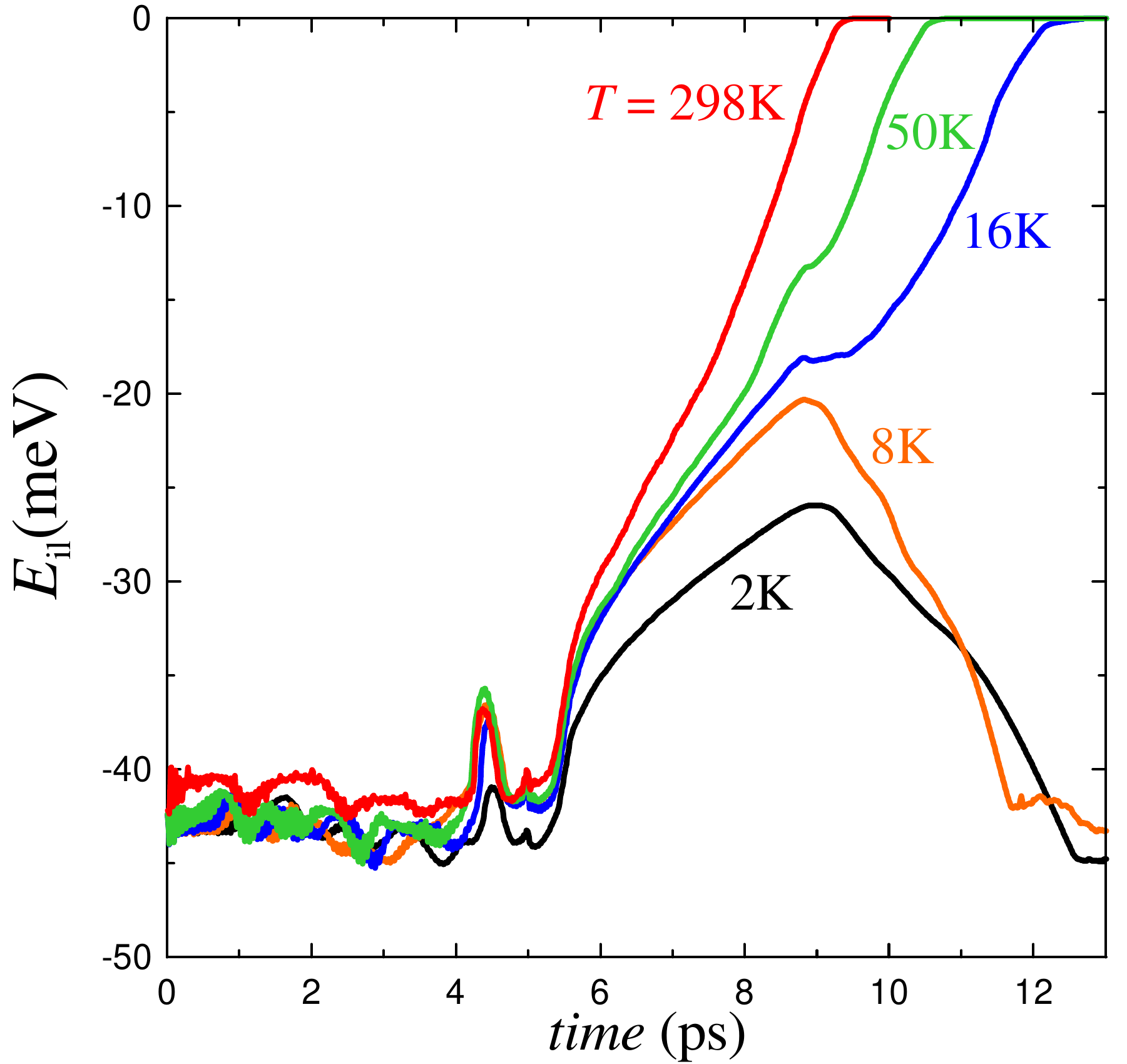}}
\caption{Time dependencies of the interlayer binding energy of the upper two graphene layers for different temperatures.}
\label{fig5}
\end{figure}

Let us consider the behavior of the model at $T=298$~K in more detail (fig.~\ref{fig4} shows several snapshots of the system for this case). Following an initial slow variation of the force between the graphite substrate and the tungsten nanoasperity as the latter is being pulled toward the surface, the onset of an instability is observed, signified by a sharp increase in the attraction between the two (it corresponds to the negative value of the force). This is accompanied by a sharp decrease in the potential energy $E_{\mathrm{pot}}$ of the system (fig.~\ref{fig3}\textit{a}). The maximum attraction (point \textit{A} in fig.~\ref{fig2}\textit{a} and fig.~\ref{fig3}\textit{a}, see also fig.~\ref{fig4}\textit{a}) corresponds to a jump-to-contact (JC) phenomenon~\cite{HeoSJ2005,Landm1990} which occurs via a fast process where carbon atoms under the asperity displace toward it in a short time span of about $0.5$~ps. This phenomenon is further evidenced by time dependencies of the interlayer binding energy $E_{\mathrm{il}}$ for the upper two graphene layers shown in fig.~\ref{fig5}, where it corresponds to local maximums of $E_{\mathrm{il}}$ observed in between 4 and 5~ps. JC leads to the collision of carbon atoms with absolutely rigid nanoasperity, which causes a sharp peak in the force-displacement curve (point \textit{B} in fig.~\ref{fig2}\textit{a} and fig.~\ref{fig3}\textit{a}). Further advancement of the tip towards the sample results in an increase of the repulsion indicating the repulsive wall region and the indentation of the sample~\cite{HeoSJ2005}. Repulsion also continues to increase during the initial stage of the retraction of the tip (\textit{CD} segment of the curves) which begins after 5~ps. Moderate decay of repulsion till point \textit{E} is followed by the segment \textit{EF} where the force becomes repulsive once more and it retains the positive sign till point \textit{G} is reached. This exhibits the tendency of carbon atoms to push the tip upwards. Note, that the part of the force-versus-distance curve corresponding to the time earlier than about 8.9~ps does not unambiguously indicate the cleavage of the sample, as will be discussed later. A sudden change of repulsion to attraction after point \textit{G} in fig.~\ref{fig2}\textit{a} and fig.~\ref{fig3}\textit{a} at about 8.5~ps is indicative of the final stage of exfoliation, where forces between graphene sheets at their boundaries should be overcome. The ultimate configuration has the completely removed upper layer (fig.~\ref{fig4}\textit{d}) corresponding to zero interlayer energy in fig.~\ref{fig5}.

The behavior of the system does not change qualitatively in the range of temperatures from 298~K to 16~K, as video animations show and as can be seen from fig.~\ref{fig2}\textit{b}-\textit{c} and fig.~\ref{fig5}. Lowering the temperature leads to the decay of thermal fluctuations and it is manifested in the decrease of data scattering in the force-versus-distance curves. This also reduces the magnitudes of collisions of carbon atoms with the tip after JC and causes the reduction of the peak observed at point \textit{B} in fig.~\ref{fig2}\textit{a} and fig.~\ref{fig3}\textit{a}. One can note that in the specified temperature range the interlayer energy $E_{\mathrm{il}}$ increases during the retraction of the tip reaching zero value (fig.~\ref{fig5}) indicating the complete exfoliation of the upper layer.

\begin{figure*}[htb]
\centerline{\includegraphics[width=0.51\textwidth]{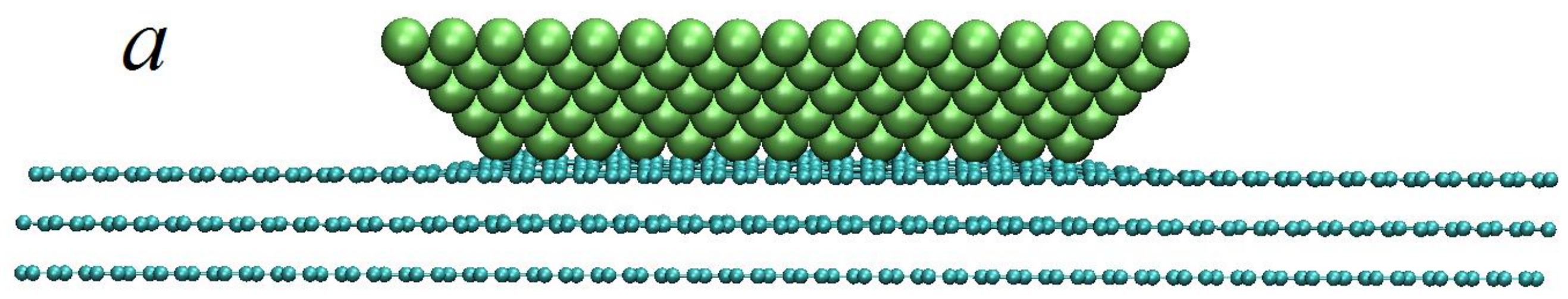}
\includegraphics[width=0.51\textwidth]{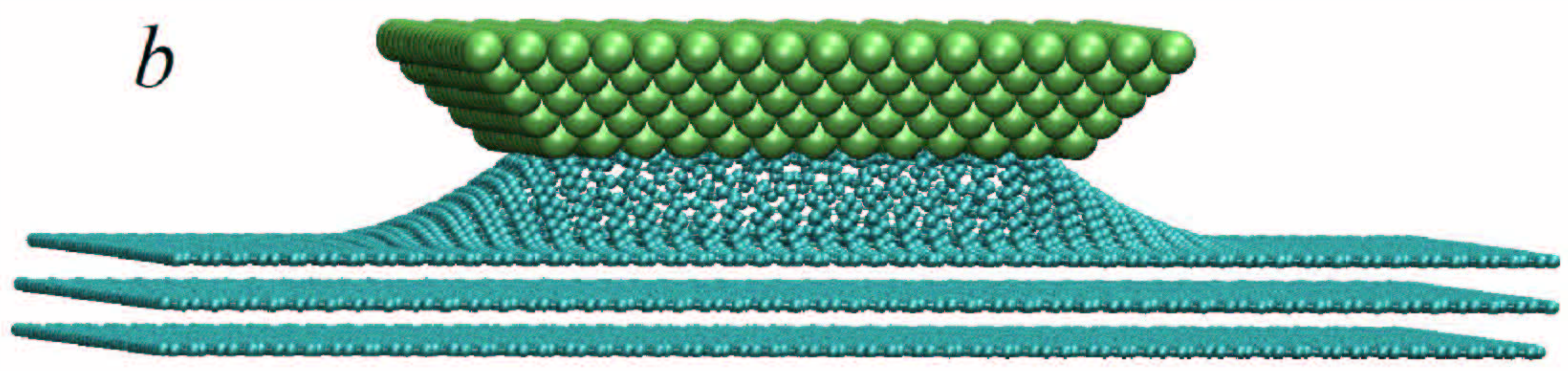}}
\centerline{\includegraphics[width=0.51\textwidth]{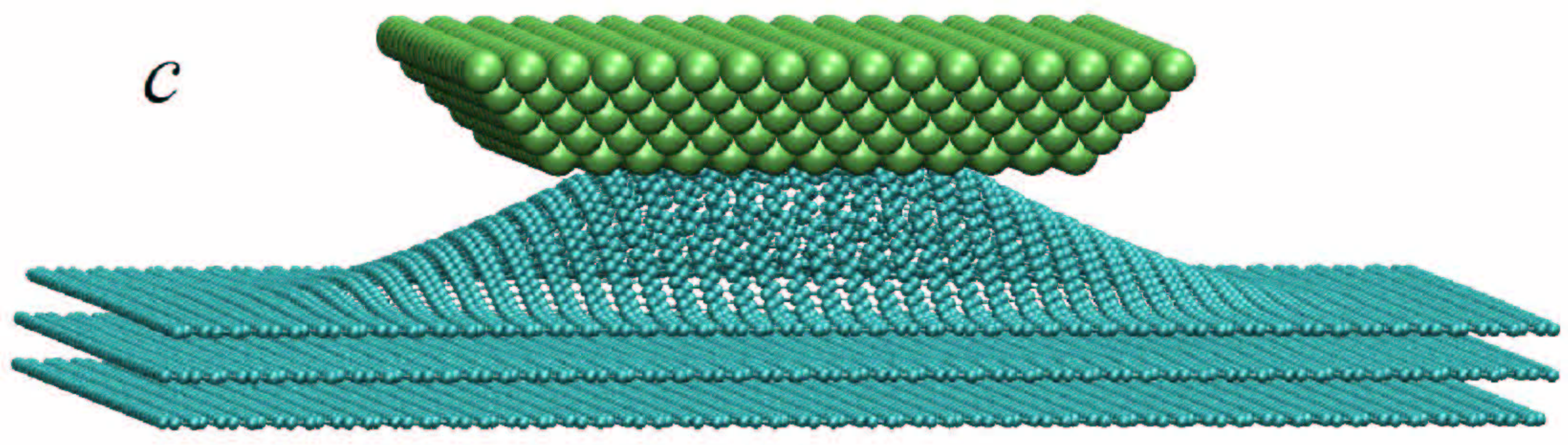}
\includegraphics[width=0.51\textwidth]{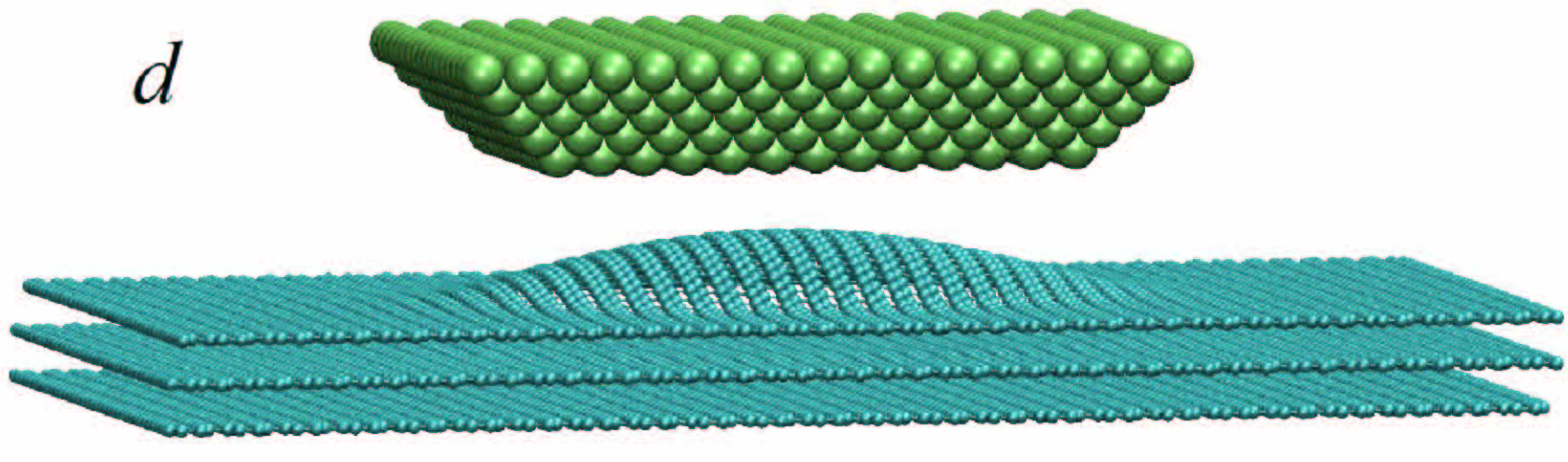}}
\caption{Snapshots of the system when $T=2$~K corresponding to the following points in fig.~\ref{fig3}\textit{b}: (\textit{a}) point \textit{A} -- 5.48~ps, (\textit{b}) \textit{B} -- 7.5~ps, (\textit{c}) \textit{C} -- 8.97~ps (maximum in fig.~\ref{fig5}), (\textit{d}) \textit{D} -- 11.5~ps, the tip ``loses'' the upper graphene layer.}
\label{fig6}
\end{figure*}

Beginning from $T=8$~K down to $T=2$~K the cleavage exhibits qualitative changes. Let us consider the case of $T=2$~K in more detail. Very small thermal fluctuations cause almost complete smoothing of the force dependencies on distance and time in fig.~\ref{fig2}\textit{d} and fig.~\ref{fig3}\textit{b} (fig.~\ref{fig6} shows several instantaneous atomic configurations for this case). When the tip is lowered toward the surface the behavior is similar to the considered one for higher temperatures with JC manifested in a sharp minimum and the following repulsive wall region. Note the absence of the peak attributed to collisions of carbon atoms with the tip. The withdrawal part of the curves is completely different from the considered above. During the retraction the force acting on the tip is mainly attractive indicating the tendency of carbon atoms to pull the tip downwards in opposite direction to the tip movement. After reaching a sharp minimum (point \textit{A} in fig.~\ref{fig3}\textit{b}) corresponding to the beginning of the pickup of carbon atoms by the tip the force remains almost constant (till point \textit{B} in fig.~\ref{fig3}\textit{b}) after which its magnitude begins to decay reaching zero value at 11.8~ps. Time dependency of $E_{\mathrm{il}}$ in fig.~\ref{fig5} exhibits maximum at 8.97~ps after which it gains the initial value in contrast to higher temperatures where absolute value of $E_{\mathrm{il}}$ decreased to zero. These facts and video animations obtained during the simulation (see also fig.~\ref{fig6}\textit{d}) clearly indicate that the tip ``loses'' carbon atoms, which return to the equilibrium positions of the upper layer, and the exfoliation does not occur.

\subsection{Exfoliation using LJP}
\label{LJP}

\begin{figure*}[htb]
\centerline{\includegraphics[width=0.51\textwidth]{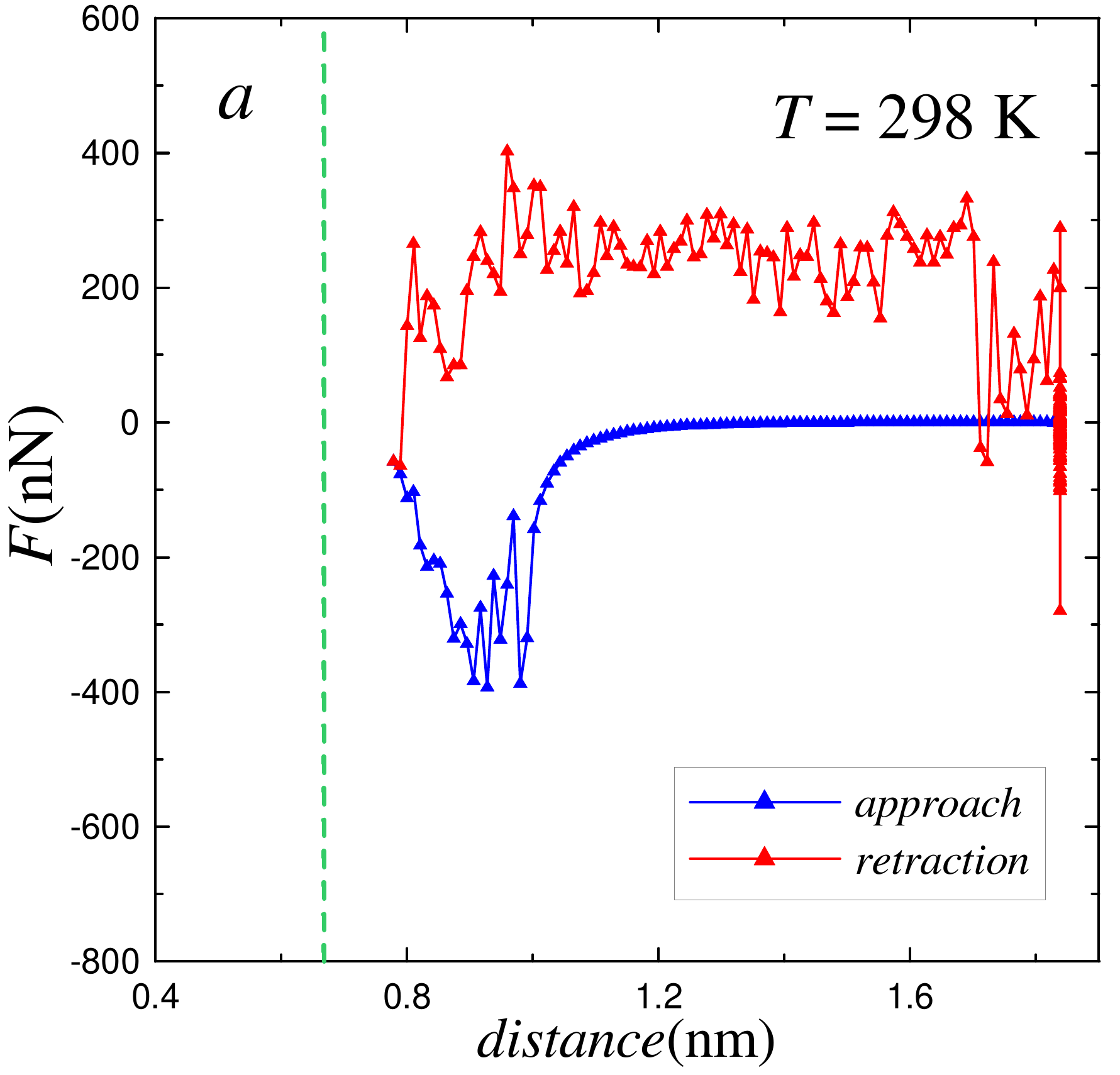}
\includegraphics[width=0.51\textwidth]{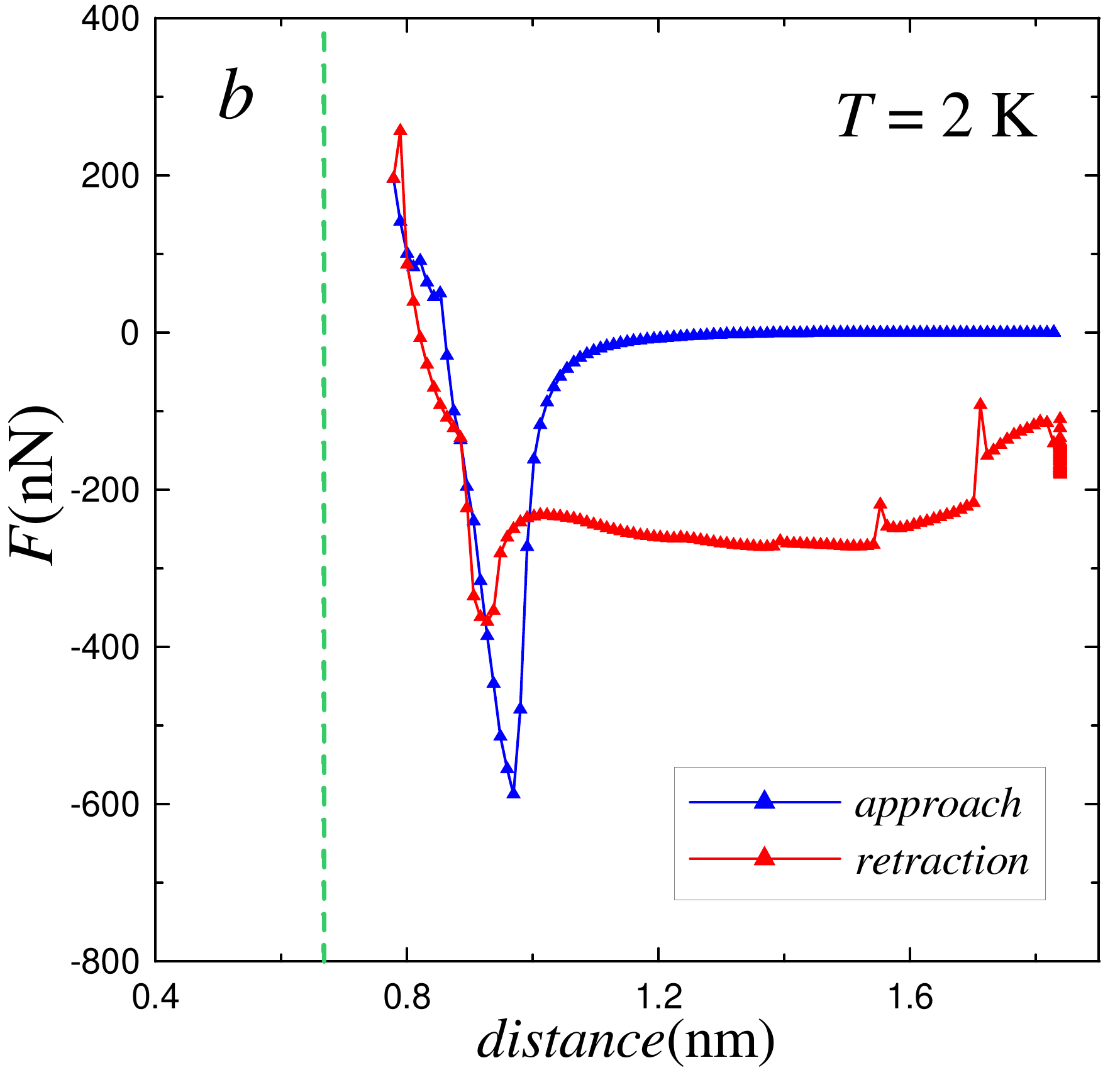}}
\caption{Force-versus-distance curves obtained with LJ interlayer potential for the maximum (\textit{a}) and minimum (\textit{b}) temperatures used in the simulations.}
\label{fig7}
\end{figure*}

\begin{figure}[!]
\centerline{\includegraphics[width=0.52\textwidth]{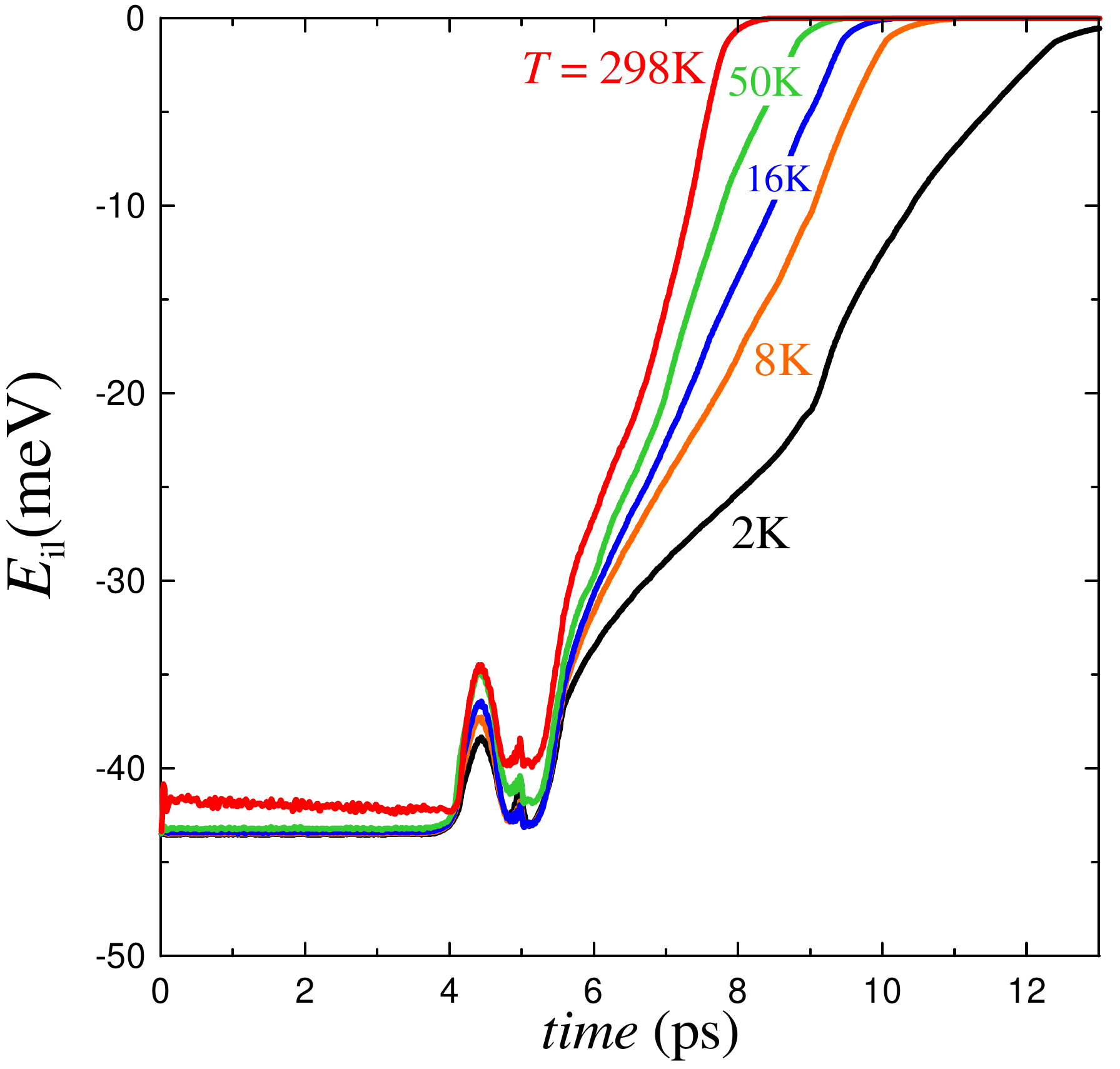}}
\caption{Time dependencies of the binding energy of the upper two graphene layers measured in the simulations employing LJP under different temperatures.}
\label{fig8}
\end{figure}

\begin{figure}[!]
\centerline{\includegraphics[width=0.52\textwidth]{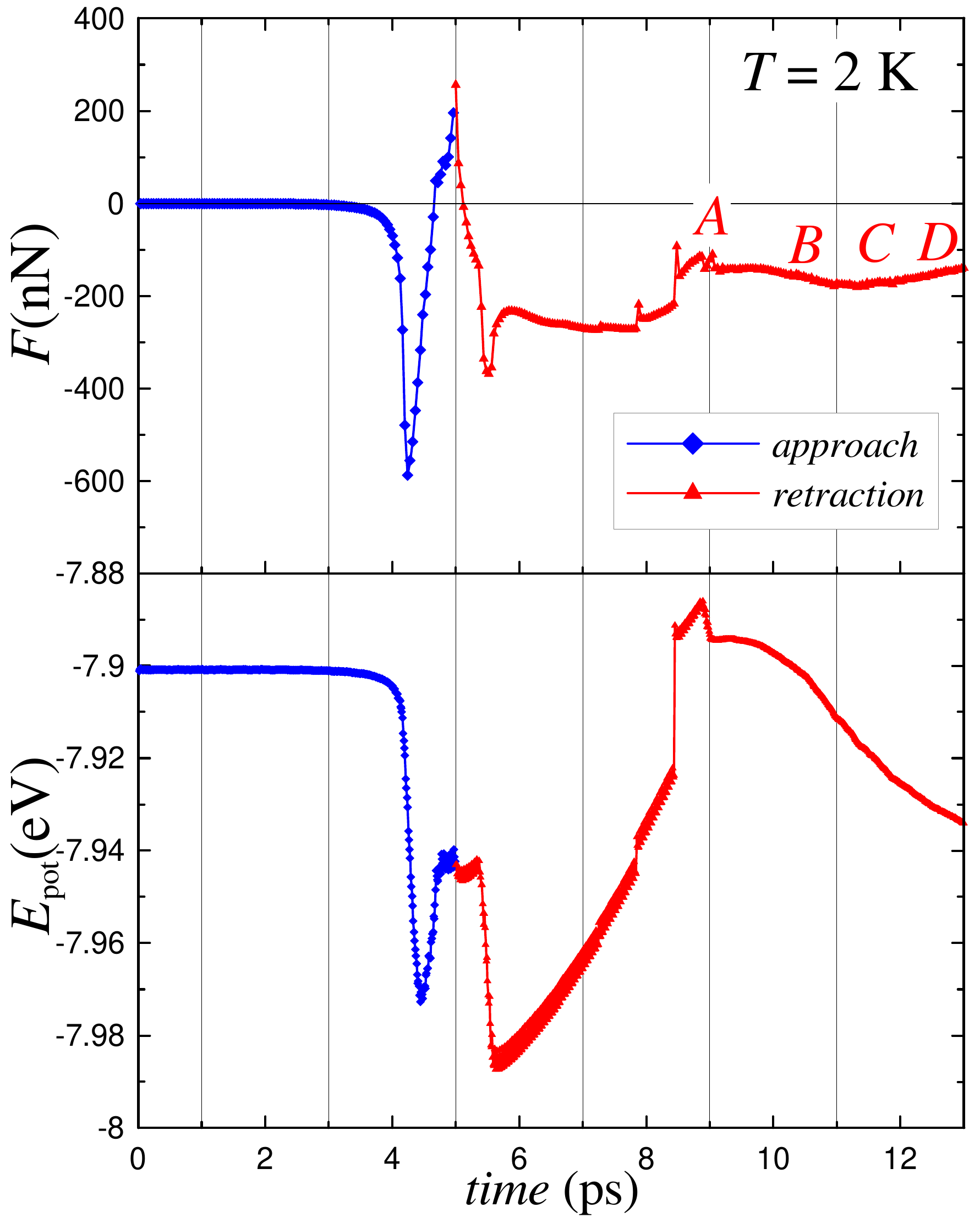}}
\caption{Time dependencies of the normal force acting on the tip and the potential energy of the system (per atom) when LJP is used for interlayer cohesion at $T=2$~K.}
\label{fig9}
\end{figure}

\begin{figure*}[htb]
\centerline{\includegraphics[width=0.51\textwidth]{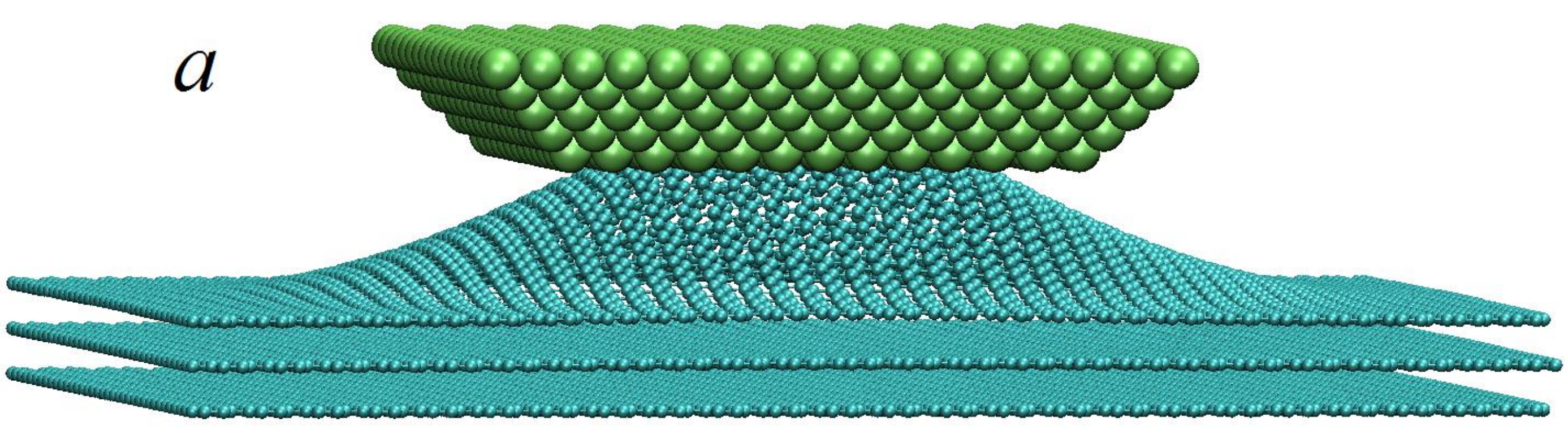}
\includegraphics[width=0.51\textwidth]{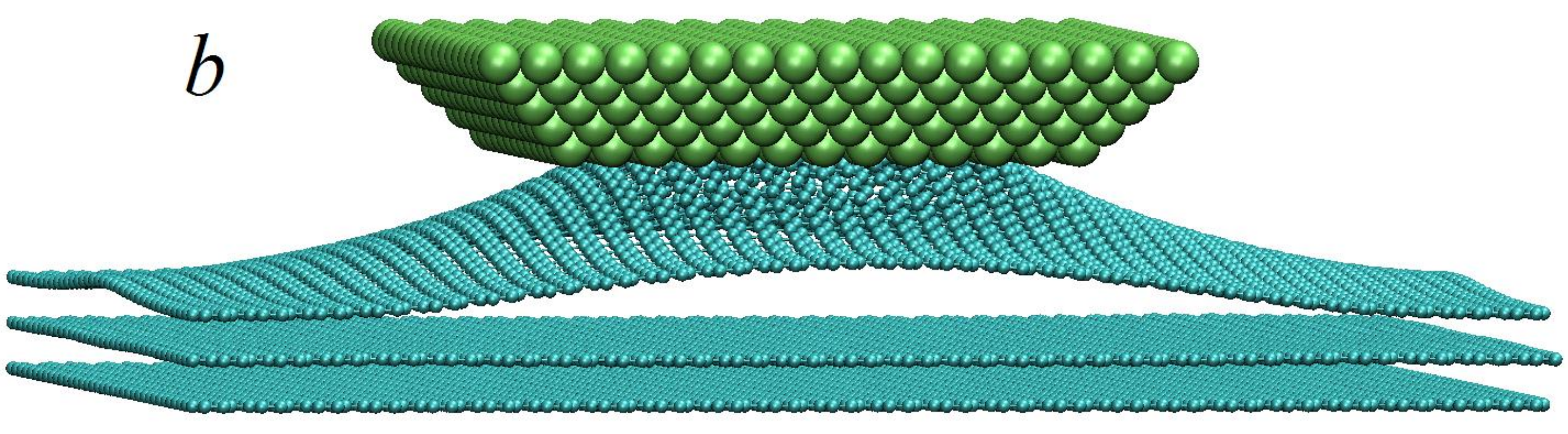}}
\centerline{\includegraphics[width=0.51\textwidth]{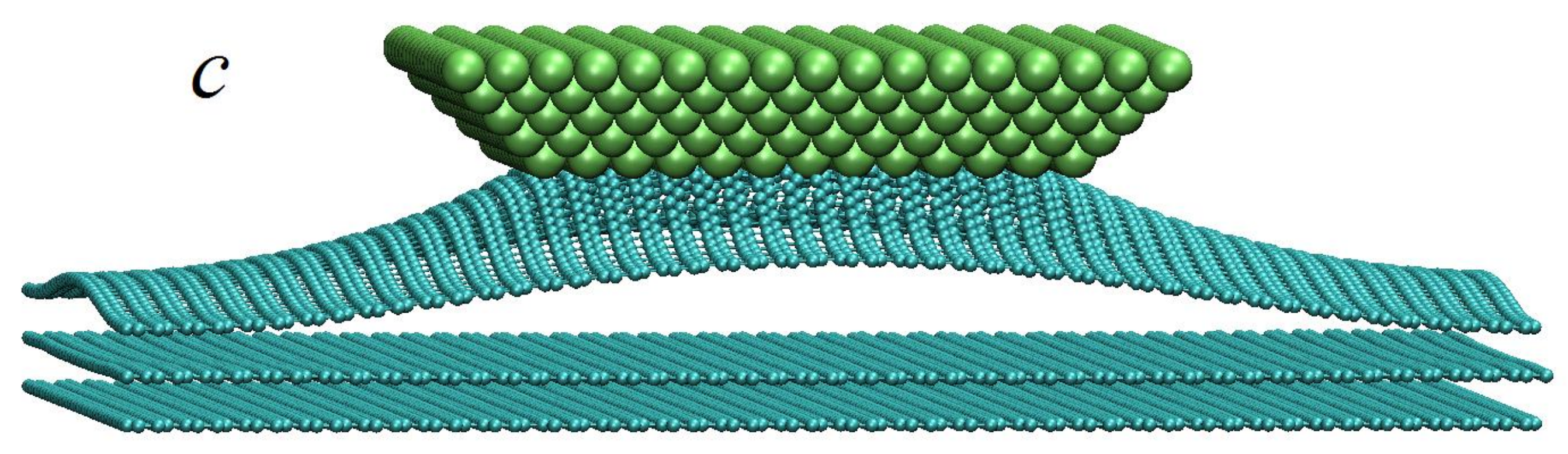}
\includegraphics[width=0.51\textwidth]{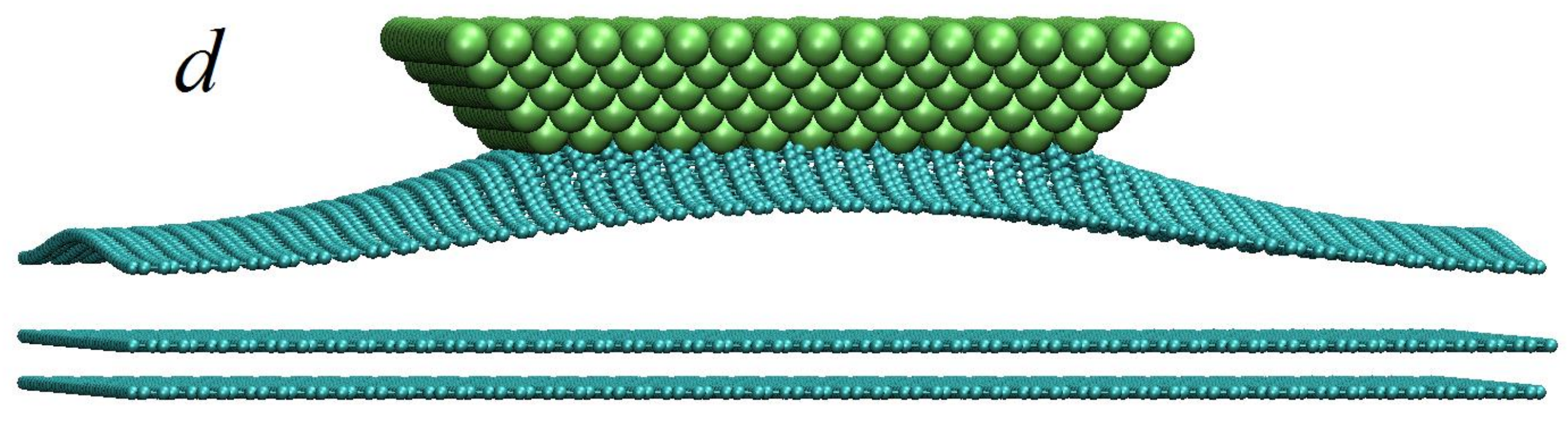}}
\caption{Snapshots of the system obtained for LJP at $T=2$~K corresponding to the following points in fig.~\ref{fig9}: (\textit{a}) point \textit{A} -- 8.97~ps, (\textit{b}) \textit{B} -- 10.5~ps, (\textit{c}) \textit{C} -- 11.5~ps, (\textit{d}) \textit{D} -- 13~ps, the tip almost completely isolates the upper graphene layer.}
\label{fig10}
\end{figure*}

Main results obtained when LJP is used for the coupling of graphitic interfaces are presented in fig.~\ref{fig7}~--~\ref{fig10}. For temperatures above 16~K force-displacement curves (see fig.~\ref{fig7}\textit{a} for $T=298$~K) and time dependencies of the force and potential energies are qualitatively similar to the obtained with RDP (with smaller magnitude of data scattering). As video animation sequences and fig.~\ref{fig8} also suggest, in the mentioned temperature range the exfoliation of the sample takes place.

For temperatures beginning from 8~K down to 2~K the behavior of $F$ and $E_{\mathrm{pot}}$ during the time span before 8.9~ps is also qualitatively similar to the considered for RDP. For later times force dependencies differ qualitatively from RDP, which is manifested in the nonzero value of $F$ that slowly changes in time (see fig.~\ref{fig9} for $T=2$~K). The interlayer energy ultimately approaches zero value in the whole range of the temperatures (fig.~\ref{fig8}), suggesting that the upper graphene layer is also cleaved at low temperatures (as can be seen in fig.~\ref{fig10} for $T=2$~K) and the tip does not ``lose'' the upper graphene layer.

\section{Discussion}
\label{discussion}

\subsection{Qualitative analysis}
\label{quality}

\begin{figure}[htb]
\centerline{\includegraphics[width=0.52\textwidth]{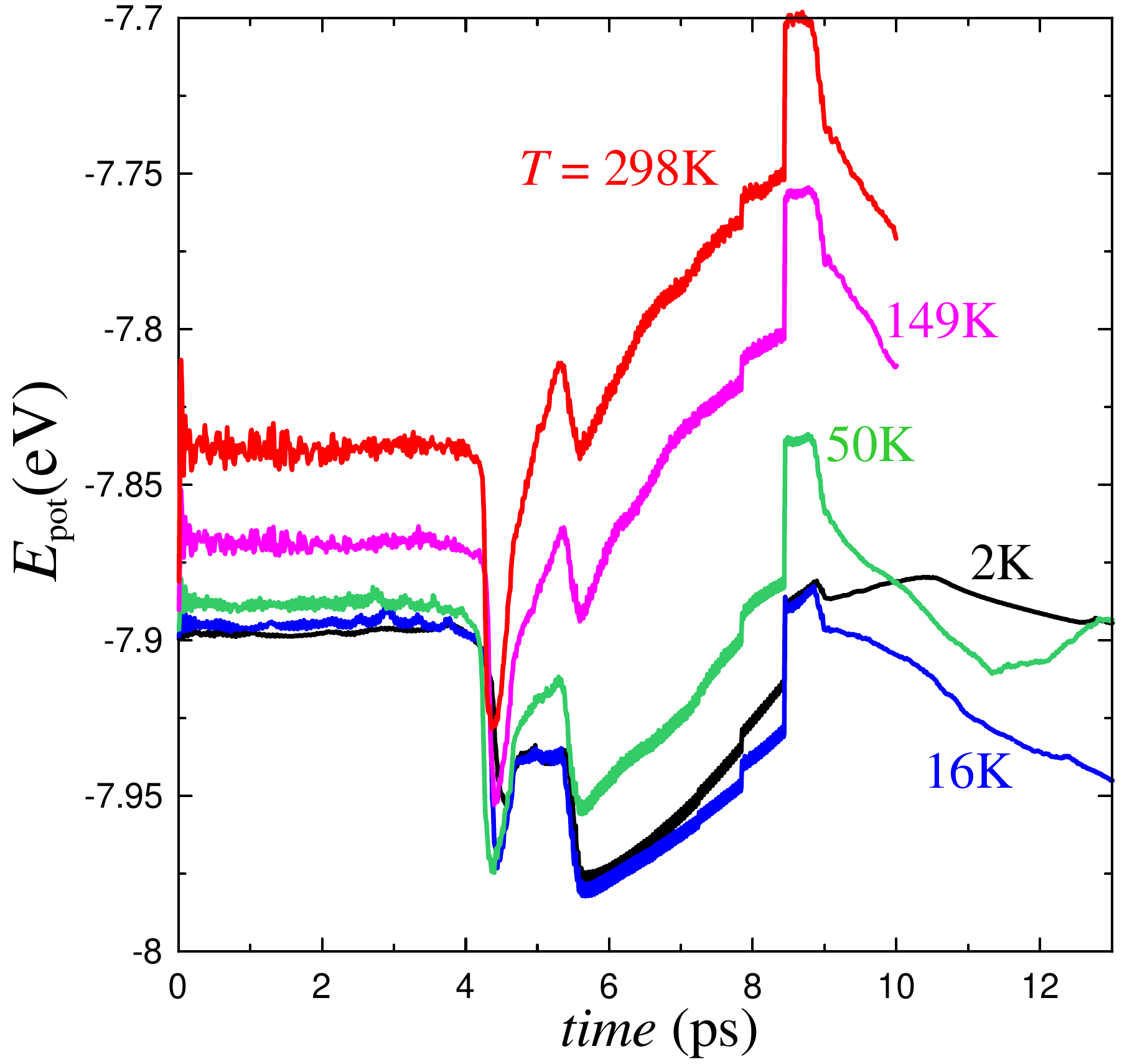}}
\caption{Time dependencies of the potential energy of the system for different temperatures obtained using RDP.}
\label{fig11}
\end{figure}

\begin{figure}[htb]
\centerline{\includegraphics[width=0.52\textwidth]{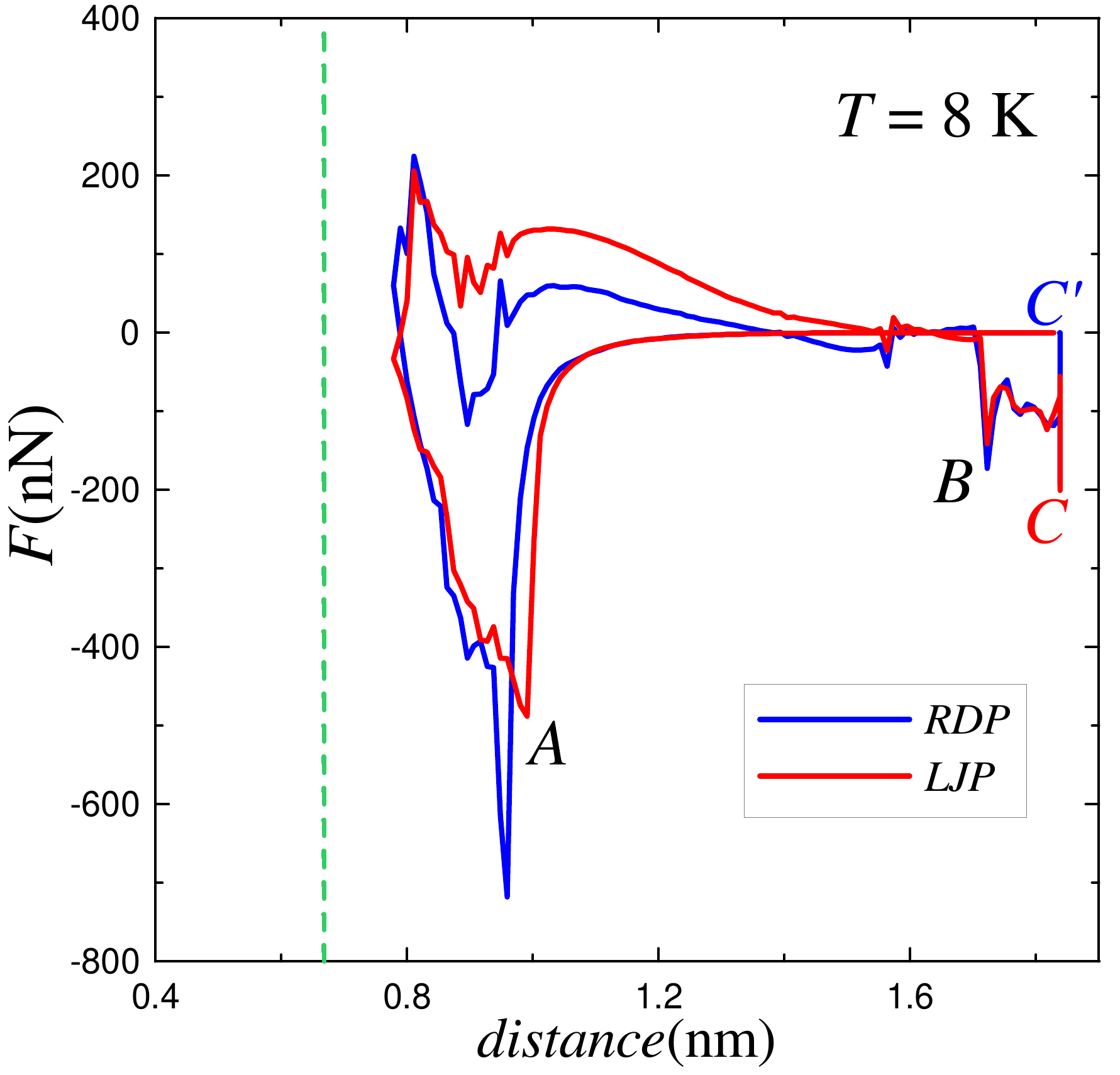}}
\caption{Force-versus-distance curves obtained using LJP and RDP at $T=8$~K.}
\label{fig12}
\end{figure}

Before revealing the mechanisms leading to qualitative differences in the exfoliation obtained using the two interlayer potentials, let us more deeply analyze the potential energy $E_{\mathrm{pot}}$ of the system and the force dependencies which also may be valuable for elucidating the interlayer behavior. As the binding energy $E_{\mathrm{il}}$ in graphite is smaller than $E_{\mathrm{pot}}$ by more than two orders of magnitude, its contribution to the total potential energy is negligible. Changes in $E_{\mathrm{pot}}$ are mainly defined by structural transformations in layers, by tip--sample interactions and by mechanical stresses occurring in the upper layer. Time dependencies of $E_{\mathrm{pot}}$ for different temperatures obtained using RDP are summarized in fig.~\ref{fig11} (for LJP they are qualitatively similar).
As was mentioned in subsection~\ref{RDP}, the sharp decrease of $E_{\mathrm{pot}}$ after 4~ps is due to JC phenomenon (see, for example, point \textit{A} in fig.~\ref{fig2}\textit{a}, fig.~\ref{fig3}\textit{a}, fig.~\ref{fig12}). It is followed by the fast increase, reflecting the compression of the sample, and then diminishing of $E_{\mathrm{pot}}$ is observed till the local minimum is reached at about 5.6~ps. This corresponds to the initial part of retraction of the nanoasperity beginning after 5~ps. The steady increase of $E_{\mathrm{pot}}$ in the time interval from 5.6~ps till about 8.5~ps reflects the deformation of the upper graphene layer during the withdrawal of the tip.
Note that for temperatures higher than 25~K (is not shown in fig.~\ref{fig11}) the value of $E_{\mathrm{pot}}$ at the moment immediately before the sharp increase of the potential energy at 8.5~ps is greater than its initial value before the JC. This is caused by the instability due to the use of the old bond order term in REBO mentioned in section~\ref{model}. Accordingly to video animations, this results in the rearrangements of atoms in some parts of the upper layer into configurations different from the honeycomb one, leading to the formation of point defects or even of large areas with disordered structure. This process is more intensive under higher $T$ and may influence the exfoliation. However, for temperatures lower than 25~K the honeycomb structure of the upper layer is preserved and this is reflected in the values of $E_{\mathrm{pot}}$. The sudden jump of $E_{\mathrm{pot}}$ occurs after 8.5~ps and it corresponds to the moment when the interlayer binding at the boundaries of the layer begins to play an important role. Its onset does not depend on $T$ indicating that this jump is defined by the geometry of the model and the tip--sample interaction, but not by the potential functions used for the interactions between carbon atoms. As video animations show, at this moment the tip abruptly ``loses'' some part of carbon atoms which results in sharp increase in $E_{\mathrm{pot}}$, and the corresponding force dependencies show the decrease of magnitude of $F$ (see, for example, point \textit{G} in fig.~\ref{fig3}\textit{a}, jump in fig.~\ref{fig3}\textit{b} before point \textit{C}, and point \textit{B} in fig.~\ref{fig12}). This is a crucial moment for exfoliation and it is important that in our model this moment occurs immediately before the retraction stops at 8.96~ps. If the tip had been withdrawn further, it would have ``lost'' atoms also for LJP at lower temperatures.

The above consideration indicates that force-displacement curves do not unambiguously reflect the possibility of the exfoliation for times less than 8.96~ps, as at this period they mainly reflect the state of carbon atoms under the tip but not of the whole layer. This explains the qualitatively similar behavior of force curves for the two potentials (see also fig.~\ref{fig12}). For times greater than 8.96~ps cleavage corresponds to zero force (when no atoms interact with the tip, point \textit{C}' in fig.~\ref{fig12}) which is observed for RDP. Comparing the dependencies of $E_{\mathrm{pot}}$ for $T=16$~K and $T=2$~K in fig.~\ref{fig11} and corresponding curves for $F$ in fig.~\ref{fig2}\textit{c} and fig.~\ref{fig2}\textit{d}, respectively, may appear to be inconsistent at a first glance, as the force curves are very different in contrast to $E_{\mathrm{pot}}$. However, accordingly to the above discussion, the discrepancy may be ascribed to the different behavior of carbon atoms under the tip during retraction. At the higher temperature they tend to move upwards after each withdrawal step, thus pushing the tip in this direction, but for the lower temperature due to smaller mobility they pull the asperity in the opposite direction. It should be noted that decrease of $E_{\mathrm{pot}}$ with diminishing the temperature may be explained using the technique applied below in subsection \ref{phenomenology} for interlayer energy, but it is out the scope of the current investigation.

Time dependencies of the interlayer binding energy $E_{\mathrm{il}}$ of the upper two graphene layers (fig.~\ref{fig5} and fig.~\ref{fig8}) provide the means to unambiguously treat the interlayer processes. The apparent interaction of the upper layer with the tip is manifested in the maximum after about 4~ps due to JC, although small but nonzero value of the force occurs after about 2~ps. As fig.~\ref{fig5} and fig.~\ref{fig8} suggest, $E_{\mathrm{il}}$ diminishes with the decrease of $T$. Averaging of $E_{\mathrm{il}}$ and of the interlayer distance $d$ during the time span from 1~ps (after equilibration) till 2~ps (when the tip does not interact with the sample) shows that for LJP mentioned tendency is observed in the whole temperature range, suggesting that thermal expansion of the interlayer spacing occurs with increase of $T$. For RDP this behavior is also observed for temperatures higher than 16~K. However, $E_{\mathrm{il}}$ and $d$ do not decrease for RDP when $T$ diminishes from 8~K to 2~K and even the opposite trend is observed. This may be ascribed to a rather small averaging time interval not enough to obtain the true average values, as for RDP much higher fluctuations of $E_{\mathrm{il}}$ are observed. On the other hand, these results may point out that although at high temperatures thermal expansion greatly facilitates cleavage, at low $T$ its significance diminishes and another contribution plays a crucial role.

\begin{figure}[htb]
\centerline{\includegraphics[width=0.52\textwidth]{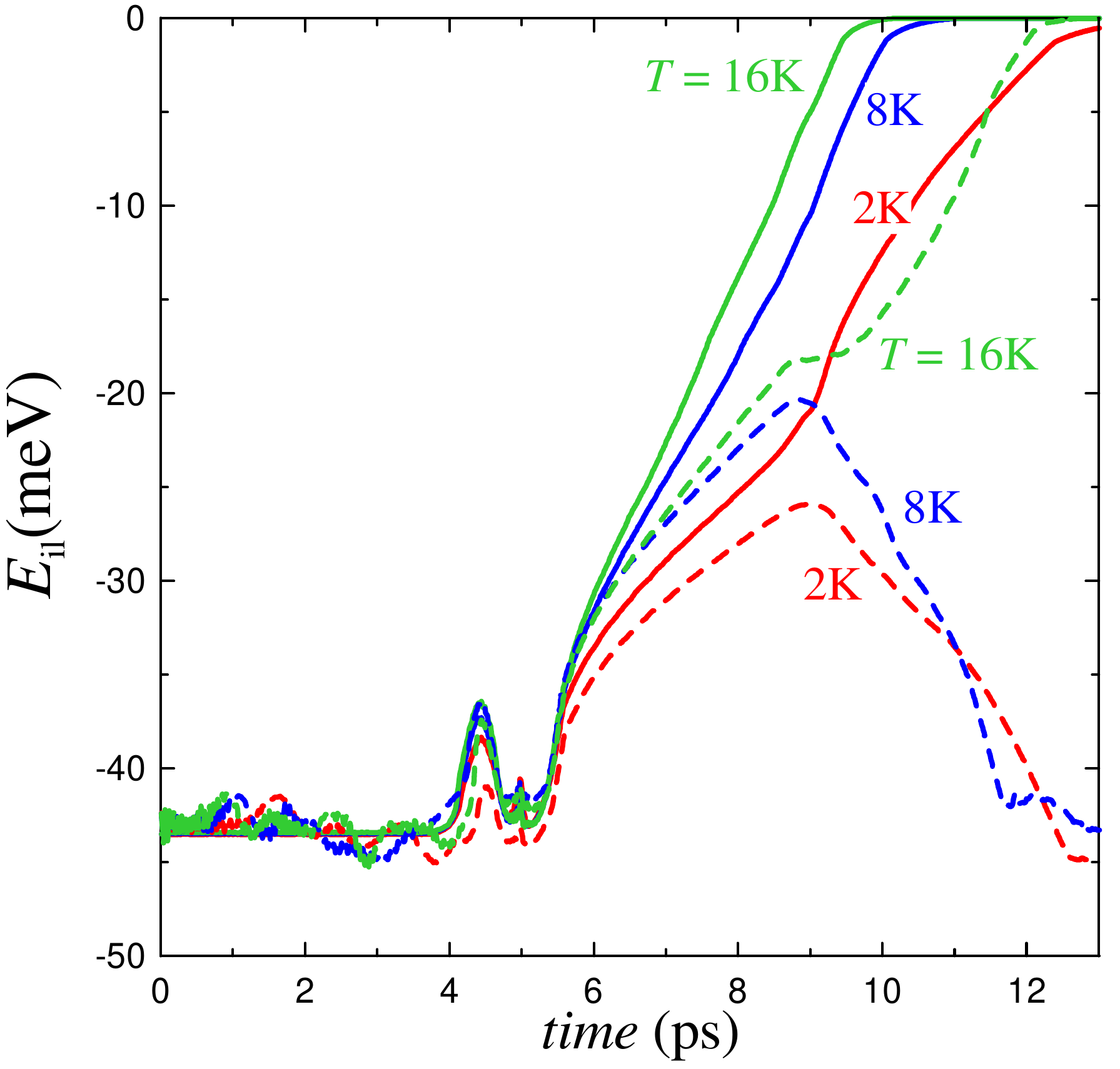}}
\caption{Time dependencies of the binding energy of the upper two graphene layers obtained using RDP (dashed lines) and LJP (solid lines) at low temperatures.}
\label{fig13}
\end{figure}

The withdrawal parts of curves for $E_{\mathrm{il}}$ at high temperatures are monotonic and qualitatively similar. They may suggest that defects in the upper layer and thermal expansion facile the cleavage process. For LJP with decrease of $T$ the slope of the curves diminishes, indicating the slowing of the exfoliation, but the form of the dependency is preserved down to 2~K (see also fig.~\ref{fig13}). In contrast, for RDP lowering $T$ leads to qualitative changes. At about 50~K the curve becomes nonmonotonic at about 8.5~ps, which corresponds to a moment of a sharp jump in $E_{\mathrm{pot}}$ (see the above discussion). Further decrease of $T$ transforms bending of the curve into a plateau and ultimately into a maximum. These results clearly suggest that for RDP a potential barrier exists, which should be overcome at the moment after about 8.5~ps to provide cleavage of the sample. The barrier is not apparent at high temperatures but manifests itself with decrease of $T$. In our model, beginning from 8~K the magnitudes of the tip--sample interaction and of the stresses in the upper layer are not enough to overcome the barrier and thus the exfoliation does not occur.

\subsection{Phenomenology}
\label{phenomenology}

Now we make some analytical estimates which may help to conceive the appearance of the energy barrier for RDP and to explain the observed differences for the two potentials. Note, however, that we do not pretend to obtain accurate quantitative results and the main aim is to reveal the main trends.

\begin{figure*}[htb]
\centerline{\includegraphics[width=0.48\textwidth]{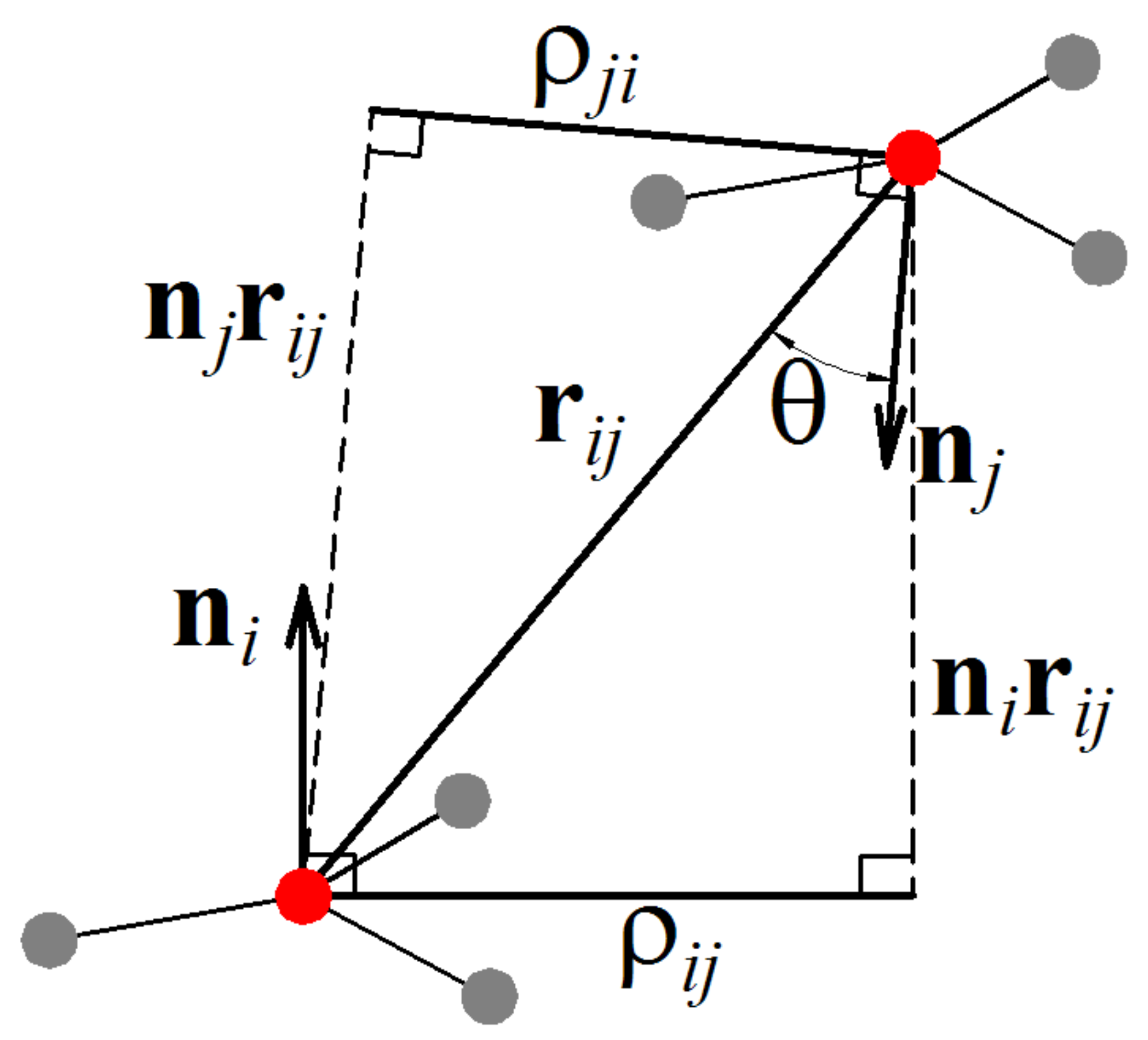}
\includegraphics[width=0.52\textwidth]{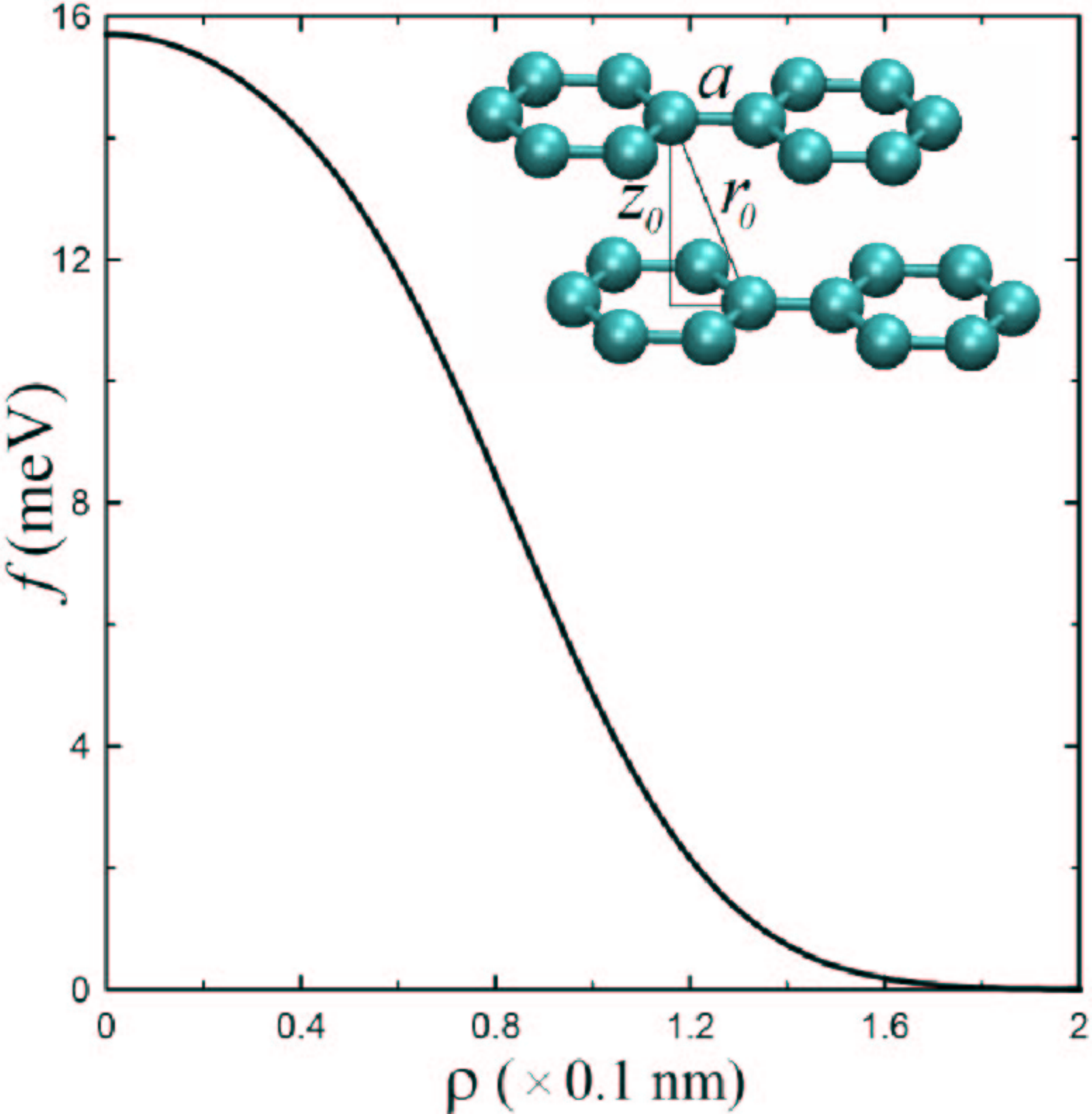}}
\caption{(\textit{a}) Geometrical illustration of the quantities that appear in eq.~(\ref{eq4-f}),~(\ref{eq5-rho}), and (\textit{b}) rapidly decaying $f(\rho)$ dependence, $f\approx0$ for $\rho>\rho_{cr}\approx2$~\AA.}
\label{fig14}
\end{figure*}

At first, let us analyze the behavior of the RDP under different temperatures. The main feature of this potential is function $f$ (see eq.~(\ref{eq2-kolmogorov})) of the following form~\cite{Kolmo2005}:
\begin{equation}
\label{eq4-f}
f(\rho)=
\left[
C_{0} + C_{2}\left(\frac{\rho}{\delta}\right)^2
      + C_{4}\left(\frac{\rho}{\delta}\right)^4
\right]
\exp{\left[-\left(\frac{\rho}{\delta}\right)^2\right]}.
\end{equation}
Here $C_{0}=15.71$, $C_{2}=12.29$, $C_{4}=4.933$ are measured in meV and $\delta=0.578$~\AA. Quantity $f$ reflects the directionality of the overlap of $\pi$ orbitals and makes the dominant contribution to the repulsive part of the RDP. It rapidly decays with the transverse distance $\rho$ (fig.~\ref{fig14}), which is defined using local normal $\mathbf{n}_{k}$ in the vicinity of an atom $k$:
\begin{eqnarray}
\label{eq5-rho}
\rho_{ij}^2=r_{ij}^2-(\mathbf{n}_{i}\mathbf{r}_{ij})^2,
\rho_{ji}^2=r_{ij}^2-(\mathbf{n}_{j}\mathbf{r}_{ij})^2.
\end{eqnarray}

For $\alpha$ form of graphite atoms in distinct layers can appear one under another, or an atom in one layer can be located under the center of a honeycomb of another layer. We consider only the latter situation, for the former the trends should be qualitatively similar. It is sufficient to study the contributions to $f$ only of the first neighbors as the transverse distance $\rho$ to the second and farther neighbors is more than 2~\AA,~and thus their contribution to $f$ is negligible (see fig.~\ref{fig14}\textit{b}).
\begin{figure}[htb]
\centerline{\includegraphics[width=0.52\textwidth]{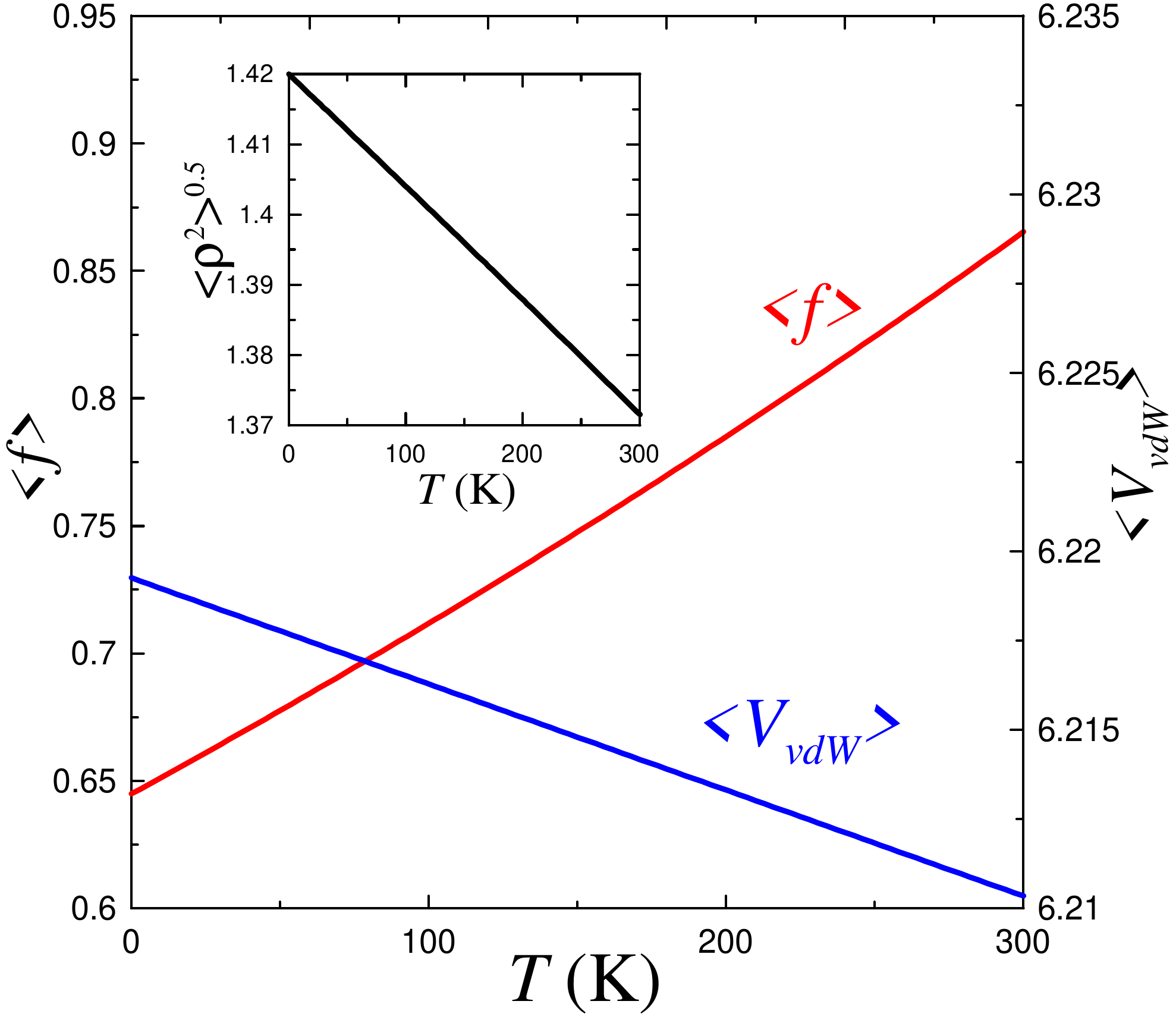}}
\caption{Temperature dependencies of $\langle f\rangle$, $\langle V_{vdW}\rangle$ and $\langle\rho^2\rangle$. Energies are measured in meV, $\sqrt{\langle\rho^2\rangle}$ is in \AA.}
\label{fig15}
\end{figure}
For static layers the distance between the mentioned two atoms from distinct layers is $r_{0}=\sqrt{z_{0}^2+a^2}$, where $a=1.42$~\AA~is the interatomic distance in a graphite layer (see inset in fig.~\ref{fig14}\textit{b}). In dynamic layers radius-vector connecting such two atoms can be written as follows:
\begin{equation}
\label{eq6-r}
\mathbf{r}(t)=\mathbf{r}_{0}+\mathbf{u}(t),
\end{equation}
where $\mathbf{u}(t)$ is a stochastic quantity that changes in time due to thermal fluctuations. Thermal motion of atoms also leads to fluctuations of normals, that are used to determine the orientation of $\pi$ orbitals. Therefore, the angle $\theta$ between the vectors $\mathbf{r}$ and $\mathbf{n}$ (fig.~\ref{fig14}\textit{a}) also fluctuates and we can write
\begin{equation}
\label{eq7-theta}
\theta(t)=\theta_{0}+\widetilde{\theta}(t),
\end{equation}
here $\theta_{0}=\arcsin(a/r_{0})$ corresponds to static layers and $\widetilde{\theta}(t)$ changes in a stochastic manner. Substituting (\ref{eq6-r}), (\ref{eq7-theta}) in (\ref{eq5-rho}), and applying averaging by sufficiently long time interval gives the following expression for a mean square of $\rho$ (we omit indices for simplicity):
\begin{eqnarray}
\label{eq8-rho2}
\left\langle\rho^2\right\rangle=\left\langle r^2\right\rangle-\left\langle r^2\cos^2\theta\right\rangle=
\left\langle r_{0}^2+u^2+2\mathbf{r}_{0}\mathbf{u}\right\rangle
\nonumber\\
-\left\langle\left(r_{0}^2+u^2+2\mathbf{r}_{0}\mathbf{u}\right)\cos^2(\theta_{0}+
\widetilde{\theta})\right\rangle.
\end{eqnarray}
Since the normal of an atom is defined using the nearest neighbors in the layer, and vector $\mathbf{u}$ pertains to atoms in different layers, these quantities can be assumed to fluctuate independently. Thus, averaging the expressions containing $\mathbf{u}$ and $\widetilde{\theta}$ can be performed separately. With this in mind, applying standard trigonometric identity and taking into account a small magnitude of $\widetilde{\theta}$ and that $\left\langle\mathbf{r}_{0}\mathbf{u}\right\rangle=0$, one obtains
\begin{eqnarray}
\label{eq9-rho3}
\left\langle\rho^2\right\rangle=
r_{0}^2+\left\langle u^2\right\rangle-
r_{0}^2\left\langle\left(\cos^2\theta_{0}-\widetilde{\theta}\sin2\theta_{0}+
\widetilde{\theta}^2\sin^2\theta_{0}\right)\right\rangle
\nonumber\\
-\left\langle u^2\left(\cos^2\theta_{0}-\widetilde{\theta}\sin2\theta_{0}+
\widetilde{\theta}^2\sin^2\theta_{0}\right)\right\rangle.
\end{eqnarray}
Employing the equality $\langle\widetilde{\theta}\rangle=0$ we have:
\begin{eqnarray}
\label{eq10-rho-ms}
\left\langle\rho^2\right\rangle=\left(r_{0}^2+
\left\langle u^{2}\right\rangle -
r_{0}^2\langle\widetilde{\theta}^{2}\rangle-
\left\langle u^{2}\right\rangle\langle\widetilde{\theta}^{2}\rangle\right)\sin^2\theta_{0}.
\end{eqnarray}
From the thermodynamic relation $C\left\langle u^{2}\right\rangle/2= 3k_{\mathrm{B}}T/2$ follows that $\left\langle u^{2}\right\rangle=\alpha T$, where $\alpha=3k_{\mathrm{B}}/C$, $k_{\mathrm{B}}$ is the Boltzmann constant and $C$ is an effective spring constant. Substituting values of carbon atomic mass $M=1.994\cdot10^{-26}$~kg and of frequency $\omega=10^{14}$~rad$/$s in $C=M\omega^2$, one obtains $C=199.4$~N$/$m and $\alpha=2.1\cdot10^{-5}$~\AA$^2/$K. Analogously assuming that $\langle\widetilde{\theta}^{2}\rangle=\beta T$ and $\sqrt{\langle\widetilde{\theta}^{2}\rangle}=15^{\circ}$ at 300~K, we have $\beta=2.25\cdot10^{-4}$~rad$^2/$K. Eq.~(\ref{eq10-rho-ms}) takes the form
\begin{eqnarray}
\label{eq11-rho-T}
\langle\rho^2\rangle=a^2+
(\alpha T -
\beta r_{0}^2T-
\alpha\beta T^2
)a^2/r_{0}^2.
\end{eqnarray}
Inset in fig.~\ref{fig15} plots the dependency $\sqrt{\langle \rho^2\rangle}(T)$. As can be seen, it decays with $T$.

Further step is to express analytically $\langle f\rangle$ through $\langle \rho^2\rangle$. One way to do this is to expand $\exp{\left[-\left(\rho/\delta\right)^2\right]}$ into series by powers of $\rho/\delta$ near zero point and then substitute expression from eq.~(\ref{eq11-rho-T}). However, since $\rho/\delta$ is not small this leads to alternating series with increasing terms, and therefore we cannot terminate the series at finite number of terms. Direct integration for averaging also does not allow expressing analytically $\langle f\rangle$ through $\langle \rho^2\rangle$ and the numerical techniques should be involved. To avoid this, we simply substitute $\langle \rho^2\rangle$ in eq.~(\ref{eq4-f}) to obtain very rough estimate of the dependence $\langle f\rangle(T)$. The result is shown in fig.~\ref{fig15}.

For vdW interaction neighbors farther than the first one should be taken into account to obtain accurate energy values. This is the reason for relatively large cutoff distance used in the simulations which covers about 6 or 7 neighbors. Nevertheless, the major contribution makes the first neighbor and for estimative aims it is sufficient to consider it. It can be shown that with accuracy to moments higher than the second the expression $\langle r^6\rangle\approx\langle r^2\rangle^3$ is true. Using eq.~(\ref{eq6-r}), we obtain for the temperature dependence of magnitude of vdW interaction from eq.~(\ref{eq2-kolmogorov}):
\begin{eqnarray}
\label{eq12-vdW}
\langle V_{vdW}\rangle=\frac{Az_{0}^6}{\langle(\mathbf{r}_{0}+\mathbf{u})^6\rangle}\approx
\frac{Az_{0}^6}{\langle r_{0}^2+u^2\rangle^3}=
\frac{Az_{0}^6}{\langle r_{0}^2+\alpha T\rangle^3}.
\end{eqnarray}
The results are summarized in fig.~\ref{fig15}. As can be seen, $f$ increases with temperature, that corresponds to greater repulsion due to the interlayer wave-function overlap, and vdW attraction decreases, indicating thermal expansion of the interlayer separation. It can be noted that the growth rate $\mathrm{d}f/\mathrm{d}T$ is by about an order of magnitude larger than the rate of vdW decaying with $T$.

\begin{figure*}[htb]
\centerline{\includegraphics[width=1\textwidth]{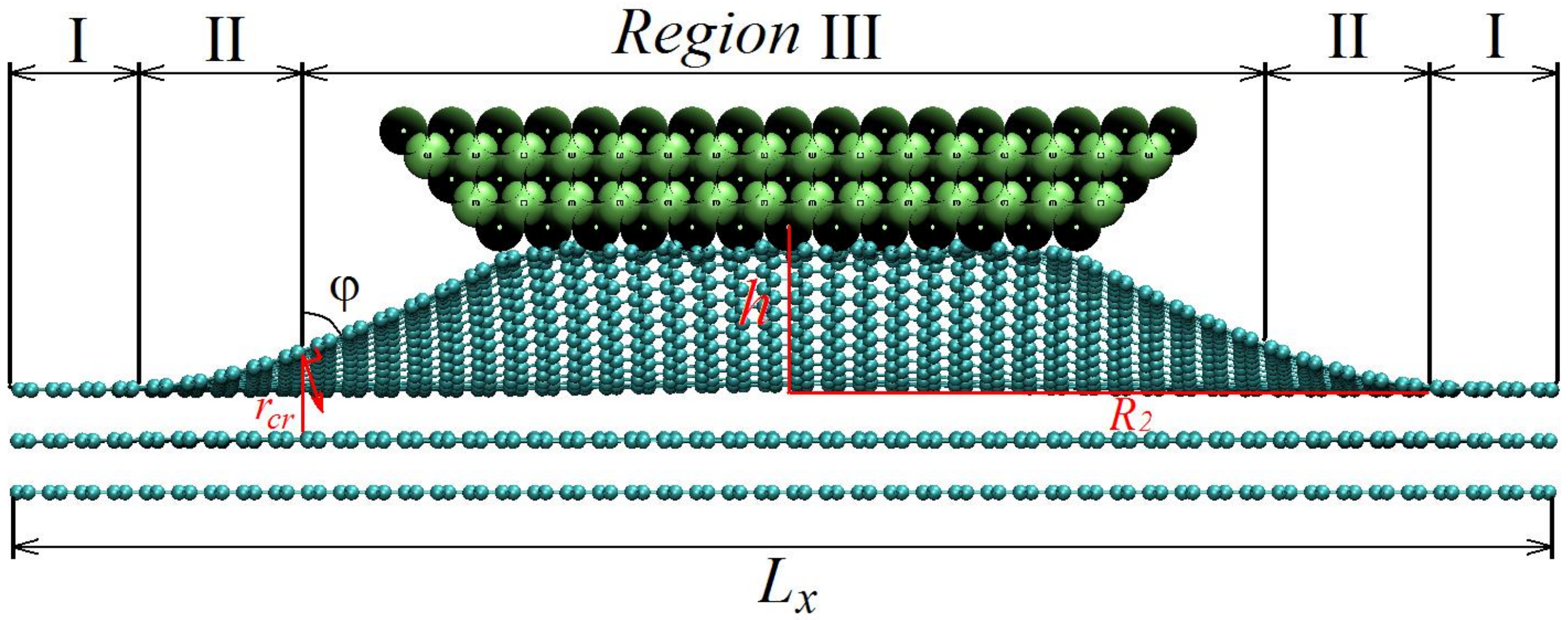}}
\caption{Sectional view of the system when $T=8$~K at 8.6~ps illustrating the three regions with different contributions to the interlayer binding.}
\label{fig16}
\end{figure*}

Let us analyze the formation of the potential barrier. It is defined by the sizes of the contact surface of the tip $a_{x}$ and $a_{y}$, the dimensions of the sample $L_{x}$, $L_{y}$, the ultimate tip height and the rate of its retraction $v$, by stiffness of the upper layer and by the energy of the tip--sample interactions $\varepsilon_{WC}$. We consider the situation which is observed in the simulations for low temperatures, when for the given $\varepsilon_{WC}$ the rate $v$ has such a value that the upper layer is neither completely cleaved nor is ``lost'' when the tip reaches the height $h$ (above the equilibrium position of the upper layer, see fig.~\ref{fig16}) at about 8.6~ps corresponding to jumps in $E_{\mathrm{pot}}$ immediately before the tip stops. Three regions with different contributions to the interlayer interaction can be marked out (fig.~\ref{fig16}).
\begin{enumerate}
\item Region I -- atoms are not effected by the tip. Here both contributions -- repulsive from the $\pi$ orbital overlap $V_{\pi}^{I}$ and attractive vdW $V_{vdW}^{I}$ are presented (these quantities stand for average values of the interlayer energy).
\item Region II -- both contributions $V_{\pi}^{II}$ and $V_{vdW}^{II}$ exist but the magnitude of $V_{\pi}^{II}$ decreases to 0 when approaching to region III.
\item Region III -- only vdW attraction $V_{vdW}^{III}$ is presented.
\end{enumerate}
For the fixed values of $v$, $\varepsilon_{WC}$, $h$ and of the interlayer potential the dimensions of these zones are defined by the geometry of the system and the stiffness of the layer. We denote by $R_{2}$ the distance from the center of symmetry of the layers to the boundary between the regions I and II, and by $R_{3}$ the one to the boundary between regions II and III (boundaries are assumed to be the circles on the surface of the layer). From fig.~\ref{fig16} we can conclude that
\begin{eqnarray}
\label{eq13-R2}
R_{2}=a_{x}/2+h\tan\varphi,
\end{eqnarray}
here we use the fact $a_{x}=a_{y}$ in our model. The quantity $R_{3}$ corresponds to some critical distance $r_{cr}$ between the layers, where $\rho=\rho_{cr}=2$~\AA~and therefore $f\approx0$. As the angle between the normal to the upper layer and $\mathbf{r}_{cr}$ is $\pi/2-\varphi$, hence $\rho^2=r_{cr}^2-r_{cr}^2\cos^2(\pi/2-\varphi)$ and ultimately $r_{cr}=\rho_{cr}/\cos\varphi\approx4.7$~\AA~for $\varphi=65^{\circ}$ (the value chosen only for estimative purpose). From fig.~\ref{fig16} one can note, that
\begin{eqnarray}
\label{eq14-R3}
R_{3}=R_{2}-(r_{cr}-z_{0})\tan\varphi.
\end{eqnarray}

Having defined the dimensions of the regions we can calculate the interlayer energy for the considered time moment. Denoting the number of atoms per unit surface area in the layer by $n$ and assuming the absence of defects (which is the case for low temperatures in the simulations), the numbers of atoms $N_{I}$, $N_{II}$ $N_{III}$ in each region are defined as follows
\begin{eqnarray}
\label{eq15-N}
N_{I}=n(L_{x}L_{y}-\pi R_{2}^2),
\nonumber
N_{II}=n\pi\frac{R_{2}^2-R_{3}^2}{\sin^2\varphi},
\nonumber\\
N_{III}\approx n(a_{x}a_{y}+\pi\frac{R_{3}^2-0.25a_{x}^2}{\sin^2\varphi}).
\end{eqnarray}
For the binding energy of the upper two layers we have the following expression
\begin{eqnarray}
\label{eq16-eil}
E_{il}=N_{I}(V_{\pi}^{I}+V_{vdW}^{I})+N_{II}(V_{\pi}^{II}+V_{vdW}^{II})
+N_{III}V_{vdW}^{III}.
\end{eqnarray}
Eq.~(\ref{eq16-eil}) presents a wrapped form of the approximation for the potential barrier that should be overcome to provide the cleavage of the layer. We analyze this expression qualitatively. As was shown above, under the lowest temperature the repulsive contributions from the orbital overlap $V_{\pi}^{I}$, $V_{\pi}^{II}$ have the smallest possible value, and vdW attraction between the layers is the largest. Therefore, in the competition of the three types of interactions: vdW tip--sample interaction and $\pi$ orbital overlap that tend to separate the layers and vdW attraction between the layers, wins the latter and the cleavage does not take place. This situation for the chosen parameters of our model remains up to 8~K, although the interlayer barrier considerably diminishes (fig.~\ref{fig5}). Beginning from 16~K due to fast growth of $f$ with $T$ the two interactions tending to isolate the layer begin to prevail over vdW interlayer attraction (which has been diminished due to thermal expansion), the magnitude of the barrier reaches the value that can be overcome and the exfoliation occurs. The above discussion shows that it is this anisotropic orbital overlap contribution that plays an important role in the behavior of the model due to its fast growth with $T$. The pairwise interaction gives minor contribution as the result of the weaker temperature dependence.

It should be noted that another combination of the parameters of the model can result in different scenario from the observed above. For example, for much larger sample the cleavage could not have been occurred, since the first region would have much larger size. Stiffness of the layers also plays an important role as it defines the angle $\varphi$ (when the other parameters are fixed) and hence the dimensions of the marked regions. Lower stiffness leads to smaller $\varphi$ and enlarges the first zone, thus worsen the conditions for cleavage. The use of tips with reduced $a_{x}$ and $a_{y}$ will also lead to decreased the third and the second regions and hence to the probable absence of exfoliation. This firmly suggests that in simulations where another in-plane potential is employed or in experiments the observed processes would take place in distinct temperature range. There also might be the need to adjust other parameters to obtain the results described in the current work.

\section{Conclusion}
\label{concl}
Computer experiments presented in the paper reveal the influence of the temperature on the exfoliation of a graphitic sample for two different types of the interlayer potential. The main result of the simulations is that the inclusion of the anisotropic contribution accounting for $\pi$ orbital overlap into the interlayer potential can qualitatively change the kinetics of cleavage under low temperatures, although the potentials can give almost the same interlayer cohesive energy. In our model this is manifested in the absence of the exfoliation below 8~K when RDP is used, while for LJP the upper graphene layer is isolated in the whole range of the considered temperatures. Some analytical estimates have been carried out that qualitatively explain the behavior observed in the simulations and define the contributions of different geometrical parameters of the system to the considered processes. The results obtained indicate the need of the experimental verification of the role of orbital overlap in the interlayer cohesion of graphite and the correctness of its description by the RDP. Our model provides a sketch for the possible experimental setup, which can be based on the electrostatic exfoliation technique presented in Ref.~\cite{Liang2009} as it allows adjusting the magnitude of the tip--sample interactions.



\textbf{Acknowledgements}

We acknowledge the use of a Windows beowulf cluster at Informatics department of Sumy State University.

\end{document}